\documentclass[11pt,onecolumn,aps,superscriptaddress, amsmath,amssymb,nofootinbib]{revtex4}   
\pdfoutput=1
\usepackage{latexsym}
\usepackage{graphics}

\usepackage{amsmath}    
\usepackage{graphicx}   
\usepackage{verbatim}   
\usepackage{color}      
\usepackage{subfigure}  
\usepackage{hyperref}   

\newcommand{\bmat}{\left(\begin{array}}
\newcommand{\emat}{\end{array}\right)}
\newcommand{\beq}{\begin{equation}}
\newcommand{\eeq}{\end{equation}}

\newcommand{\drawsquare}[2]{\hbox{%
\rule{#2pt}{#1pt}\hskip-#2pt
\rule{#1pt}{#2pt}\hskip-#1pt
\rule[#1pt]{#1pt}{#2pt}}\rule[#1pt]{#2pt}{#2pt}\hskip-#2pt
\rule{#2pt}{#1pt}}

\newcommand{\fund}{\raisebox{-.5pt}{\drawsquare{6.5}{0.4}}}
\newcommand{\Ysymm}{\raisebox{-.5pt}{\drawsquare{6.5}{0.4}}\hskip-0.4pt%
        \raisebox{-.5pt}{\drawsquare{6.5}{0.4}}}
\newcommand{\Yasymm}{\raisebox{-3.5pt}{\drawsquare{6.5}{0.4}}\hskip-6.9pt%
        \raisebox{3pt}{\drawsquare{6.5}{0.4}}}
\newcommand{\antifund}{\overline{\fund}}

\def\yzero{\smash{\hbox{$y\kern-4pt\raise1pt\hbox{${}^\circ$}$}}}

\def\-{\hphantom{-}}
\def\ov{\overline}
\def\s2{\frac{1}{\sqrt2}}

\def\beq{\begin{equation}}
\def\eeq{\end{equation}}
\def\beqa{\begin{eqnarray}}
\def\eeqa{\end{eqnarray}}

\def\IF{\relax{\rm I\kern-.18em F}}
\def\II{\relax{\rm I\kern-.18em I}}
\def\IP{\relax{\rm I\kern-.18em P}}

\def\Dsl{\,\raise.15ex\hbox{/}\mkern-13.5mu D} 

\def\IC{\bf C}
\def\IZ{\bf Z}
\def\IT{\bf T}
\def\z2z2{$\IC^3/(\IZ_2\times\IZ_2)$}

\def\s{\sigma}

\def\z{\zeta}

\def\bo{{\raise-.3ex\hbox{\large$\Box$}}}               
\def\face{{\raise.2ex\hbox{$\displaystyle \bigodot$}\mskip-2.2mu \llap {$\ddot
        \smile$}}}                                      

\def\leftrightarrowfill{$\mathsurround=0pt \mathord\leftarrow \mkern-6mu
        \cleaders\hbox{$\mkern-2mu \mathord- \mkern-2mu$}\hfill
        \mkern-6mu \mathord\rightarrow$}       
\def\dvec#1{\vbox{\ialign{##\crcr
        \leftrightarrowfill\crcr\noalign{\kern-1pt\nointerlineskip}
        $\hfil\displaystyle{#1}\hfil$\crcr}}}           

\def\beq{\begin{equation}}
\def\eeq{\end{equation}}

\def\beqx{\begin{displaymath}}
\def\eeqx{\end{displaymath}}

\def\beqa{\begin{eqnarray}}
\def\eeqa{\end{eqnarray}}

\begin{document}

\title{
\normalsize \mbox{ }\hspace{\fill}
\begin{minipage}{12 cm}
\end{minipage}\\[5ex]
{\large\bf  Revisiting the Supersymmetric Pati-Salam Models \\
                 from Intersecting D6-branes 
\\[1ex]}}
\author{Tianjun Li \footnote{E-mail: \texttt{tli@itp.ac.cn}}} 
\affiliation{CAS Key Laboratory of Theoretical Physics, Institute of Theoretical Physics,\\
	Chinese Academy of Sciences, Beijing 100190, P. R. China}
\affiliation{School of Physical Sciences, University of Chinese Academy of Sciences,\\
	No.19A Yuquan Road, Beijing 100049, P. R. China}

\author{Adeel Mansha\footnote{E-mail: \texttt{adeelmansha@itp.ac.cn}}}
\affiliation{CAS Key Laboratory of Theoretical Physics, Institute of Theoretical Physics,\\
	Chinese Academy of Sciences, Beijing 100190, P. R. China}  
\affiliation{School of Physical Sciences, University of Chinese Academy of Sciences,\\
	No.19A Yuquan Road, Beijing 100049, P. R. China}

\author{Rui Sun\footnote{E-mail: \texttt{sunrui@mpp.mpg.de}}}
\affiliation{Yau Mathematical Sciences Center, Tsinghua University, \\
	Haidian District, Beijing 100084, P. R. China}



\begin{abstract}
\medskip

Employing novel random and supervised scanning methods, we systematicly revisit the  construction of three-family $N=1$ supersymmetric Pati-Salam models in Type IIA orientifolds on $\IT^6/(\mathbb Z_2\times \mathbb Z_2)$ with intersecting D6-branes. Arising from the stacks of D6-branes with $U(n)$ gauge symmetries, the Pati-Salam gauge symmetries $SU(4)_C\times SU(2)_L \times SU(2)_R$ can be broken down to the Standard Model via D-brane splitting as well as D- and F-flatness preserving Higgs mechanism. Also, the hidden sector contains $USp(n)$ branes, which are parallel with the orientifold planes or their ${\mathbb Z_2}$ images. We find that the Type II T-duality in the previous study is not an equivalent relation in Pati-Salam model building if the model is not invariant under $SU(2)_L$ and $SU(2)_R$ exchange, and provides a way to obtain new models.
We systematically construct the new models with three families, 
which usually do not have gauge coupling unification at the string scale. 
We for the first time construct the Pati-Salam models with at least one wrapping number 
whose absolute value is larger than or equal to $5$. In particular, for one large wrapping number 
equal to $5$, we find that one kind of models carries more refined gauge couplings, and thus 
with more possibilities to have approximate gauge coupling unification. 


\end{abstract} 
\maketitle

\vskip2cm
\newpage

\section{Introduction}

The goal of string phenomenology is to construct the $N=1$ supersymmetric Standard Models (SM)
or the SM from string theories. In Type I, Type IIA and Type IIB string theories, D-branes 
as boundaries of open strings plays an important role in phenomenologically interesting model 
building~\cite{JPEW}. For the open string sectors, conformal field theory provides the consistent 
constructions of four-dimensional supersymmetric $N=1$ chiral models with non-Abelian gauge symmetry 
on Type II orientifolds. Within such framework, we obtain the chiral fermions on the worldvolume of
the D-branes which are located at orbifold singularities~\cite{ABPSS, berkooz, ShiuTye, lpt, MCJW, Ibanez, MKRR}, and/or at the intersections of 
D-branes in the internal space~\cite{bdl}, which have a T-dual description in terms 
of magnetized D-branes~\cite{bachas,urangac}.

Within the  intersecting D6-brane models on Type IIA orientifolds~\cite{bgkl, bkl, afiru}, 
many non-supersymmetric three-family
Standard-like models and grand unified models have been constructed
[12$-$25]. However, they typically suffer from the
large Planck scale corrections at the loop level, or in other words,
there exists the gauge hierarchy problem. On the other hand,
a large number of the supersymmetric three-family
Standard-like models and grand unified models have been constructed 
as well~\cite{CSU1, CSU2, CP, CPS, CLS1, CLS2, MCIP, CLW, blumrecent, Honecker,
LLG3, Cvetic:2004ui, Cvetic:2004nk, Chen:2005aba, Chen:2005mm, Chen:2005mj,
Chen:2007px, Chen:2007ms, Chen:2007zu, Chen:2008rx}, which can solve the above problem.
For a review, see Ref.~\cite{Blumenhagen:2005mu}.

In Ref.~\cite{Cvetic:2004ui}, Cveti\v c, Liu and one of us (TL) systematically constructed
the three-family $N=1$ supersymmetric Pati-Salam models from Type IIA orientifolds on
$\IT^6/(\IZ_2\times \IZ_2)$ with intersecting D6-branes where all the gauge symmetries come 
from $U(n)$ branes. The Pati-Salam gauge symmetries $SU(4)_C\times SU(2)_L \times SU(2)_R$ 
can be broken down to $SU(3)_C\times SU(2)_L \times U(1)_{B-L} \times U(1)_{I_{3R}} $ via
D6-brane splittings, and further down to the SM via four-dimensional $N=1$ supersymmetry 
preserving Higgs mechanism. Thus, it provides a road to the SM without any additional  
anomaly-free $U(1)$'s around the electroweak scale. Also, 
the hidden sector contain $USp(n)$ branes, which are
parallel with the orientifold planes or their ${\bf Z_2}$ images. These models have
at least two confining gauge groups in hidden sector, whose gaugino
condensation can in turn trigger supersymmetry breaking and (some) moduli stabilization.
In particuar, Chen, Mayes, Nanopoulos and one of us (TL) found one of these models
with a realistic phenomenology~\cite{Chen:2007px, Chen:2007zu}, 
and study its variations as well~\cite{Chen:2007ms}. Thus, we shall revisit 
such kind of three-family $N=1$ supersymmetric Pati-Salam model building in this work.

Moreover, it has been pointed out that there are a few other potentially interesting constructions 
which might lead to the SM~\cite{Cvetic:2004ui}. For example, the possible massless 
vector-like Higgs fields, which do not arise from a $N=2$ subsector, can break the Pati-Salam gauge symmetry 
down to the SM or break the $U(1)_{B-L}\times U(1)_{I_{3R}}$ down to $U(1)_Y$. However,
 because  the large wrapping numbers is required by the increased absolute values of the intersection numbers  
between $U(4)_C$ stack of D-branes and $U(2)_R$ stack or its orientifold image, it might be very difficult 
to find such models. 
Another interesting scenario is to construct the $SU(2)_L$ and/or $SU(2)_R$ gauge symmetries 
from filler branes, {\it i.e.}, $SU(2)_{L,R}=USp(2)_{L,R}$. And then 
the number of the SM Higgs doublet pairs might be decreased.
However, we do not want to construct the $SU(2)_{L,R}$ gauge symmetries from the
splittings of higher rank $USp(N)$ ($N\ge 4$) branes, which would
 lead to even number of families in general.  In such case, the absolute value for
one wrapping number of $U(4)$ branes larger than 2 cannot be avoided,
which might make the model building very difficult due to the tadpole cancellation conditions. 
Interestingly, with the better scanning method, one can definitely try to construct
 these models in the future.

Employing novel random and supervised scanning methods, we will further systematically study the three-family $N=1$ supersymmetric Pati-Salam model building
in Type IIA orientifolds on $\IT^6/(\IZ_2\times \IZ_2)$ with intersecting D6-branes in which
 the $SU(4)_C\times SU(2)_L \times SU(2)_R$ gauge symmetries arise from $U(n)$ branes. In particular, we construct the new models with large winding numbers as well, and find that the approximate gauge coupling unification can be achieved at the string scale.

The paper is organized as follows. In Section \ref{sec:orientifold} we briefly review 
the basic rules for supersymmetric intersecting D6-brane model building on Type IIA $T^6/(\IZ_2\times \IZ_2)$ orientifolds,  
the tadpole cancellation conditions, and the conditions for D6-brane configurations which
preserve four-dimensional $N=1$ supersymmetry. Also, we will briefly review the T-duality symmetries and its variations in the supersymmetric model building with intersecting D6-branes.  

In Section \ref{sec:models}, we study the supersymmetric D6-brane model building
with large winding numbers and generic T-duality in consideration. 
We point out that the Type II T-duality in Ref~\cite{Cvetic:2004ui} is not
an equivalent relation in Pati-Salam model building if the model is not invariant under $SU(2)_L$ 
and $SU(2)_R$ exchange, and provides a way to obtain the new model.
With this construction, we obtain the supersymmetric D6-brane models  with only one $USp$ group 
in the hidden sector, which have three families of the SM fermions,  
as well as satisfy the tadpole cancellation conditions and $N=1$ supersymmetry preserving conditions.
Furthermore,  we for the first time expand our investigation to the models with large wrapping number, a la $5, 6, 7, 8, 9$ and $10$, 
and obtain the approximate gauge coupling unification in these models.

In Section \ref{sec:pheno}, we discuss the phenomenological consequences of new models in different classes. 
For each class, we show the full phenomenology table for one representative. As explicit examples, 
we present the chiral spectra in the open string sector for each class of models. 
The difference of T-dual model with paralleled third two-torus in spectrum can also be found in this section.

\begin{figure}[h!]
	\centering
	\includegraphics[scale=1.2]{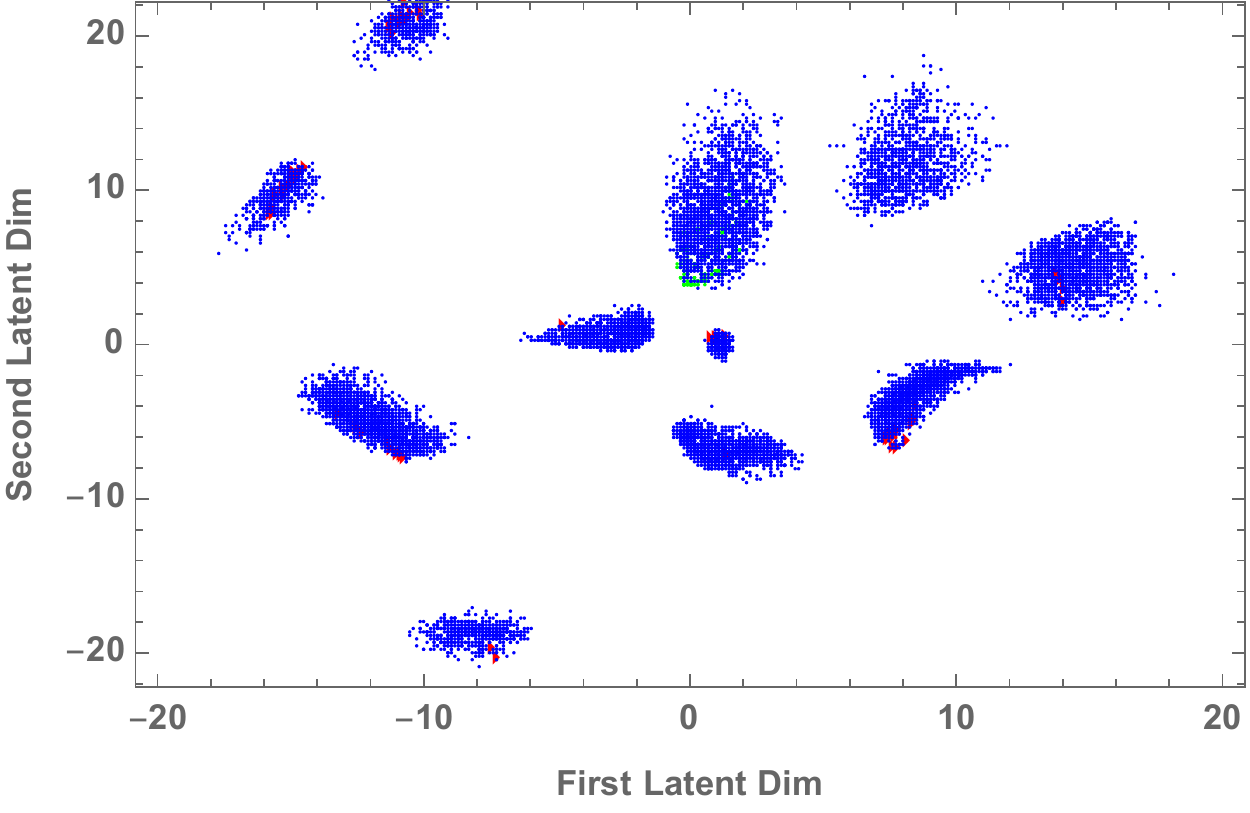}
	\caption{\label{fig:landscape2}AutoEncoder Dimension Reduction}
\end{figure}

In Section \ref{sec:plot}, we perform machine learning methods to show in Figure~\ref{fig:landscape1} and Figure~\ref{fig:landscape2} how 
the Minimal Supersymmetric SM (MSSM)-like models 
expand in our scanning according to the reduced latent dimension (which reduced from 18 wrapping numbers). 
We find that the MSSM-like models tend to gather in islands and indicates more chances to find more MSSM-like models 
in the nearby region of them.

In Section \ref{sec:conclusion}, we briefly discuss the other potentially interesting setups and conclude. 
Also, we present the D6-brane configurations and intersection numbers for supersymmetric Pati-Salam models
in the Appendix.

\section{$T^6 /(Z_2 \times Z_2)$ Orientifolds with Intersecting D6-Branes}
\label{sec:orientifold}

First, let us briefly review the basic rules to construct the supersymmetric models 
on Type IIA $T^6 /(Z_2 \times Z_2)$ orientifolds with D6-branes intersecting at generic angles, 
as well as to obtain the massless open string state spectra in Refs.~\cite{CSU2, CPS}.
In Type IIA string theory which is compactified on a $T^6 /(Z_2 \times Z_2)$ orientifold, 
we consider $T^{6}$ as a six-torus factorized as three two-tori  $T^{6} = T^{2} \times T^{2} \times T^{2}$.
The corresponding complex coordinates for the $i$-th two-torus are  $z_i$, $i=1,\; 2,\; 3$,  respectively.

The $\theta$ and $\omega$ generators for the orbifold group $Z_{2} \times Z_{2}$,
which are respectively associated with the twist vectors $(1/2,-1/2,0)$ and
$(0,1/2,-1/2)$, act on the complex coordinates $z_i$ as below
\beqa
& \theta: & (z_1,z_2,z_3) \to (-z_1,-z_2,z_3)~,~ \nonumber \\
& \omega: & (z_1,z_2,z_3) \to (z_1,-z_2,-z_3)~.~\,
\label{orbifold} \eeqa 
We implement the orientifold projection by gauging the $\Omega R$ symmetry, 
where $\Omega$ is world-sheet parity, and $R$ acts on the complex coordinates as follows
 \beqa
 R: (z_1,z_2,z_3) \to ({\ov z}_1,{\ov z}_2,{\ov
z}_3)~.~\, 
\eeqa 
Therefore, we have four kinds of orientifold 6-planes (O6-planes) respectively
for the actions of $\Omega R$, $\Omega R\theta$, $\Omega R\omega$, and $\Omega R\theta\omega$. 
In order to cancel the RR charges of O6-planes, we introduce stacks of $N_a$ D6-branes wrapping 
on the factorized three-cycles. Also,
there are two kinds of complex structures for a two-torus:
rectangular and tilted~\cite{bkl, Chen:2007zu, CSU2,CPS}, which are consistent with orientifold projection. 
The homology classes of the three cycles wrapped by the D6-brane stacks can be expressed in terms of 
 $n_a^i[a_i]+m_a^i[b_i]$ and $n_a^i[a'_i]+m_a^i[b_i]$ for the rectangular and tilted tori respectively, 
where $[a_i']=[a_i]+\frac{1}{2}[b_i]$. Thus, a generic one cycle can be labelled as $(n_a^i,l_a^i)$ in both cases, 
where in terms of the wrapping numbers $l_{a}^{i}\equiv m_{a}^{i}$ and $l_{a}^{i}\equiv 2\tilde{m}_{a}^{i}=2m_{a}^{i}+n_{a}^{i}$
for a rectangular two-torus and a tilted two-torus, respectively. 
And then $l_a^i-n_a^i$ must be even for a tilted two-torus.

Moreover, for  $a$ of $N_a$ D6-branes along the cycle $(n_a^i,l_a^i)$, we need to introduce
their $\Omega R$ images ${a'}$-stack of $N_a$ D6-branes with wrapping numbers $(n_a^i,-l_a^i)$. 
The homology three-cycles 
for  $a$ of $N_a$ D6-branes and  its orientifold image  $a'$ respectively are
\beq
[\Pi_a]=\prod_{i=1}^{3}\left(n_{a}^{i}[a_i]+2^{-\beta_i}l_{a}^{i}[b_i]\right),\;\;\;
\left[\Pi_{a'}\right]=\prod_{i=1}^{3}
\left(n_{a}^{i}[a_i]-2^{-\beta_i}l_{a}^{i}[b_i]\right)~,~\, \eeq
where $\beta_i=0$ or $\beta_i=1$ for the rectangular or tilted $i$-th two-torus, respectively.
The homology three-cycles, which are wrapped by the four O6-planes, are given by
\beq \Omega R: [\Pi_{\Omega R}]= 2^3
[a_1]\times[a_2]\times[a_3]~,~\, \eeq \beq \Omega R\omega:
[\Pi_{\Omega
R\omega}]=-2^{3-\beta_2-\beta_3}[a_1]\times[b_2]\times[b_3]~,~\,
\eeq \beq \Omega R\theta\omega: [\Pi_{\Omega
R\theta\omega}]=-2^{3-\beta_1-\beta_3}[b_1]\times[a_2]\times[b_3]~,~\,
\eeq \beq
 \Omega R\theta:  [\Pi_{\Omega
R}]=-2^{3-\beta_1-\beta_2}[b_1]\times[b_2]\times[a_3]~.~\,
\label{orienticycles} \eeq 
Thus, the intersection numbers can be expressed in terms of wrapping numbers as follows
\beq
I_{ab}=[\Pi_a][\Pi_b]=2^{-k}\prod_{i=1}^3(n_a^il_b^i-n_b^il_a^i)~,~\,
\eeq \beq
I_{ab'}=[\Pi_a]\left[\Pi_{b'}\right]=-2^{-k}\prod_{i=1}^3(n_{a}^il_b^i+n_b^il_a^i)~,~\,
\eeq \beq
I_{aa'}=[\Pi_a]\left[\Pi_{a'}\right]=-2^{3-k}\prod_{i=1}^3(n_a^il_a^i)~,~\,
\eeq \beq {I_{aO6}=[\Pi_a][\Pi_{O6}]=2^{3-k}(-l_a^1l_a^2l_a^3
+l_a^1n_a^2n_a^3+n_a^1l_a^2n_a^3+n_a^1n_a^2l_a^3)}~,~\,
\label{intersections} \eeq 
where $k=\beta_1+\beta_2+\beta_3$ is the total number of tilted two-tori, 
and $[\Pi_{O6}]=[\Pi_{\Omega R}]+[\Pi_{\Omega R\omega}]+[\Pi_{\Omega
R\theta\omega}]+[\Pi_{\Omega R\theta}]$ is the sum of four O6-plane
homology three-cycles.

The generic massless particle spectrum for intersecting D6-branes at general angles,
which is valid for both rectangular and tilted two-tori, can be expressed 
via the intersection numbers as listed in Table \ref{spectrum}.
In addition, the two main constraints on 
the four-dimensional $N=1$ supersymmetric model building from Type IIA orientifolds 
with intersecting D6-branes are: RR tadpole cancellation conditions and 
$N=1$ supersymmetry preservation in four dimensions, which are given in the
following subsections {\bf A} and {\bf B}, respectively.

\begin{table}[t] 
\caption{ 
General massless particle spectrum for intersecting D6-branes at generic angles.
In this table, the representations refer to $U(N_a/2)$,
the resulting gauge symmetry because of $Z_2\times Z_2$ orbifold projection~\cite{CSU2}.  
The chiral supermultiplets contain both scalars and fermions in such supersymmetric constructions. 
And in our convention, the positive intersection numbers give us the left-handed chiral supermultiplets.
 }
\renewcommand{\arraystretch}{1.25}
\begin{center}
\begin{tabular}{|c|c|}
\hline {\bf Sector} & \phantom{more space inside this box}{\bf
Representation}
\phantom{more space inside this box} \\
\hline\hline
$aa$   & $U(N_a/2)$ vector multiplet  \\
       & 3 adjoint chiral multiplets  \\
\hline
$ab+ba$   & $I_{ab}$ $(\fund_a,\antifund_b)$ fermions   \\
\hline
$ab'+b'a$ & $I_{ab'}$ $(\fund_a,\fund_b)$ fermions \\
\hline $aa'+a'a$ &$\frac 12 (I_{aa'} - \frac 12 I_{a,O6})\;\;
\Ysymm\;\;$ fermions \\
          & $\frac 12 (I_{aa'} + \frac 12 I_{a,O6}) \;\;
\Yasymm\;\;$ fermions \\
\hline
\end{tabular}
\end{center}
\label{spectrum}
\end{table}

\subsection{The RR Tadpole Cancellation Conditions}

The tadpole cancellation conditions directly lead to the $SU(N_a)^3$ 
cubic non-Abelian anomaly cancellation~\cite{Uranga,imr,CSU2}, 
while the cancellation of $U(1)$ mixed gauge and gravitational anomaly 
or $[SU(N_a)]^2 U(1)$ gauge anomaly can be achieved by Green-Schwarz mechanism 
mediated by untwisted RR fields~\cite{Uranga,imr,CSU2}.
The D6-branes and orientifold O6-planes, which are the sources of RR fields, 
are restricted by the Gauss law in a compact space, namely,
the sum of the RR charges of D6-branes and O6-planes must be zero due to 
the conservations of the RR field flux lines. The conditions 
for RR tadpole cancellations are given by
\begin{eqnarray}
\sum_a N_a [\Pi_a]+\sum_a N_a
\left[\Pi_{a'}\right]-4[\Pi_{O6}]=0~,~\,
\end{eqnarray}
where the last terms asise from the O6-planes, which have $-4$ RR charges 
in D6-brane charge unit. 

For simplicity, we define the following products of wrapping numbers 
\beq
\begin{array}{rrrr}
A_a \equiv -n_a^1n_a^2n_a^3, & B_a \equiv n_a^1l_a^2l_a^3,
& C_a \equiv l_a^1n_a^2l_a^3, & D_a \equiv l_a^1l_a^2n_a^3, \\
\tilde{A}_a \equiv -l_a^1l_a^2l_a^3, & \tilde{B}_a \equiv
l_a^1n_a^2n_a^3, & \tilde{C}_a \equiv n_a^1l_a^2n_a^3, &
\tilde{D}_a \equiv n_a^1n_a^2l_a^3.\,
\end{array}
\label{variables}\eeq 

To cancel the RR tadpoles, we introduce an arbitrary number 
of D6-branes wrapping cycles along the orientifold planes, 
dubbed as ``filler branes'', which contribute to the RR tadpole cacellation conditions 
while trivially satisfy the four-dimensional $N=1$ supersymmetry conditions. 
The tadpole conditions then take the form of
\begin{eqnarray}
 -2^k N^{(1)}+\sum_a N_a A_a=-2^k N^{(2)}+\sum_a N_a
B_a= \nonumber\\ -2^k N^{(3)}+\sum_a N_a C_a=-2^k N^{(4)}+\sum_a
N_a D_a=-16,\,
\end{eqnarray}
where $2 N^{(i)}$ is the number of filler branes wrapping along
the $i$-th O6-plane that is given in Table \ref{orientifold1}.
The filler branes, which give us the $USp$ group, carry the same wrapping numbers 
as one of the O6-planes as shown in Table \ref{orientifold1}. 
When the filler branes have non-zero $A$, $B$, $C$ or $D$, we refer to the $USp$ group 
as the $A$-, $B$-, $C$- or $D$-type $USp$ group, respectively. 

\renewcommand{\arraystretch}{1.4}
\begin{table}[t] 
\caption{The wrapping numbers for four O6-planes.} \vspace{0.4cm}
\begin{center}
\begin{tabular}{|c|c|c|}
\hline
  Orientifold Action & O6-Plane & $(n^1,l^1)\times (n^2,l^2)\times
(n^3,l^3)$\\
\hline
    $\Omega R$& 1 & $(2^{\beta_1},0)\times (2^{\beta_2},0)\times
(2^{\beta_3},0)$ \\
\hline
    $\Omega R\omega$& 2& $(2^{\beta_1},0)\times (0,-2^{\beta_2})\times
(0,2^{\beta_3})$ \\
\hline
    $\Omega R\theta\omega$& 3 & $(0,-2^{\beta_1})\times
(2^{\beta_2},0)\times
(0,2^{\beta_3})$ \\
\hline
    $\Omega R\theta$& 4 & $(0,-2^{\beta_1})\times (0,2^{\beta_2})\times
    (2^{\beta_3},0)$ \\
\hline
\end{tabular}
\end{center}
\label{orientifold1}
\end{table}

\subsection{Conditions for Four-Dimensional $N = 1$ Supersymmetric D6-Brane}

In four-dimensional $N=1$ supersymmetric models, $1/4$ supercharges from ten-dimensional Type I T-dual 
are required to be preserved, namely, these $1/4$ supercharges survive
the orientation projection of the intersecting D6-branes and the $Z_2\times Z_2$ orbifold projection 
on the background manifold. 
It was shown that the four-dimensional $N=1$ supersymmetry can be preserved after the orientation projection 
iff the rotation angle of any D6-brane with respect to the orientifold plane is an element of $SU(3)$~\cite{bdl}, 
or in other words, $\theta_1+\theta_2+\theta_3=0 $ mod $2\pi$, where $\theta_i$ is the angle 
between the $D6$-brane and orientifold-plane in the $i$-th two-torus. 
Because the $Z_2\times Z_2$ orbifold projection will automatically be survived for such D6-brane configuration, 
the four-dimensional $N=1$ supersymmetry conditions can be written as below~\cite{CPS}
\begin{eqnarray}
x_A\tilde{A}_a+x_B\tilde{B}_a+x_C\tilde{C}_a+x_D\tilde{D}_a=0,
\nonumber\\\nonumber \\ A_a/x_A+B_a/x_B+C_a/x_C+D_a/x_D<0,
\label{susyconditions}
\end{eqnarray} 
where $x_A=\lambda,\;
x_B=\lambda 2^{\beta_2+\beta3}/\chi_2\chi_3,\; x_C=\lambda
2^{\beta_1+\beta3}/\chi_1\chi_3,\; x_D=\lambda
2^{\beta_1+\beta2}/\chi_1\chi_2$, 
where $\chi_i=R^2_i/R^1_i$ are the complex structure moduli for the the $i$-th two-torus.  
And we introduce the positive parameter $\lambda$ to put all the variables $A,\,B,\,C,\,D$ on an equal footing. 
Based on these conditions,  we can classify all 
the possible D6-brane configurations, which preserve four-dimensional $N=1$ supersymmetry, into three types: 

(1) The filler brane which has the same wrapping numbers as one of the
O6-planes in Table \ref{orientifold1}. The gauge symmetry is  $USp$ group. 
Because one and only one of the wrapping number products $A$, $B$, $C$ and $D$ has non-zero and negative value, 
 we refer to the corresponding $USp$ group as the $A$-, $B$-, $C$- or $D$-type $USp$ group as mentioned 
in the last section.

(2) The Z-type D6-brane with one zero wrapping number. There are two negative and two zero values 
in $A$, $B$, $C$ and $D$.

(3) The NZ-type D6-brane without zero wrapping number. Among $A$, $B$, $C$ and $D$,  three of them are negative
 while one of them is positive. Based on which one is positive, we can classify the NZ-type branes into
the $A$-, $B$-, $C$- and $D$-type NZ branes. Each type has two forms of wrapping numbers defined as follows
\begin{eqnarray}
A1: (-,-)\times(+,+)\times(+,+),& A2:(-,+)\times(-,+)\times(-,+);\\
B1: (+,-)\times(+,+)\times(+,+),& B2:(+,+)\times(-,+)\times(-,+);\\
C1: (+,+)\times(+,-)\times(+,+),& C2:(-,+)\times(+,+)\times(-,+);\\
D1: (+,+)\times(+,+)\times(+,-),& D2:(-,+)\times(-,+)\times(+,+).
\end{eqnarray}
To be convenient , we shall refer the Z-type and NZ-type D6-branes to be $U$-branes in the following
since they carry $U(n)$ gauge symmetry.

\subsection{T-Duality Symmetry and its Variations}

In string theory, two theories are equivalent when T-duality can be performed to map one to the other. 
This also applies to D-brane model building when two models are related by T-duality. 
For D6-brane configurations, two models are equivalent if their three two-tori as well as
their corresponding wrapping numbers for all the D6-branes are correlated by an element of 
the permutation group $S_3$ acting on three two-tori. In addition, two D6-brane configurations are 
 equivalent if their wrapping numbers on two arbitrary two-tori have the same absolute values 
but opposite sign, while their wrapping numbers on the third two-torus are the same. In this case,
 we call it as the D6-brane Sign Equivalent Principle. As T-duality is not the key discussion point 
in our work,  we refer to Ref.~\cite{Cvetic:2004ui} for the details about how T-dualities and its variants
 perform in intersecting D6-brane model building technically.

\section{Supersymmetric Pati-Salam Model Building }
\label{sec:models}
\subsection{Construction of Supersymmetric Pati-Salam Models}

To construct the SM or SM-like models from the intersecting D6-brane scenarios.
besides the $U(3)_C$ and $U(2)_L$ gauge symmetries from stacks of branes,  
we must have at least two extra $U(1)$ gauge groups in both supersymmetric and non-supersymmetric models 
to obtain the correct quantum number for right-handed charged leptons~\cite{imr,CSU2,CPS,CP}. 
One is the lepton number symmetry $U(1)_L$, while the other is similar to the third
component of right-handed weak isospin $U(1)_{I_{3R}}$. And then the hypercharge is given by
\begin{eqnarray}
Q_Y=Q_{I_{3R}}+{{Q_B-Q_{L}}\over{2}}~,~\,
\end{eqnarray}
where $U(1)_B$ is the overall $U(1)$ of $U(3)_C$.  
In general, the $U(1)$ gauge symmetry, which comes from a non-Abelian $SU(N)$ gauge symmetry, is anomaly free and then its gauge field is massless. In our model building, $U(1)_{B-L}$ and $U(1)_{I_{3R}}$ arise
from $SU(4)_C$ and $SU(2)_R$ gauge symmetries, respectively. Thus, they are anomaly free and 
their gauge fields is massless.

If $U(1)_{I_{3R}}$ arises from the stack of D6-branes on top of orientifold~\cite{CSU2,CP},  
{\it i.e.}, from the $USp$ group,
there exist at least $8$ pairs 
of SM Higgs doublets, and two extra anomaly free $U(1)$ gauge symmetries in general. 
These $U(1)$ gauge symmetries could in principle be spontaneously broken via the Higgs mechanism 
by the scalar components of the chiral superfields whose quantum numbers are the same as
the right-handed neutrinos. However, the D-flatness conditions cannot be preserved,
 and then supersymmetry is broken. 
Thus, the scale of symmetry breaking should be around the electroweak scale. 
Moreover, we typically do not have any other candidates, which can preserve the
D-flatness and F-flatness conditions, and break
these gauge symmetries at an intermediate scale.

Therefore, similar to Ref.~\cite{Cvetic:2004ui},  we concentrate on the Pati-Salam models 
in which $U(1)_{I_{3R}}$ arises from the $U(2)_R$ symmetry.
Because it is very difficult to find the interesting models with $SU(2)_L$ 
from the D6-branes on the top of O6-plane~\cite{Cvetic:2004ui}, 
we study the supersymmetric $SU(4)_C\times SU(2)_L\times SU(2)_R$ model building
 from three stacks of D6-branes, which are not on the top of orientifold planes. 
In our model, we can break the
Pati-Salam gauge symmetry down to $SU(3)_C\times SU(2)_L\times U(1)_{B-L} \times U(1)_{I_{3R}}$ 
via D6-brane splittings, and further down to the SM gauge symmetry via Higgs
mechanism with Higgs particles  from a $N=2$ subsector~\cite{Cvetic:2004ui}.
Because we do not have any extra anomaly free U(1)
gauge symmetry around the electroweak scale, we solve a generic problem in
previous constructions~\cite{CSU2,CP}.

In short,  we introduce three stacks of D6-branes, $a$, $b$, $c$ with D6-brane numbers 8, 4, and 4,
which respectively give us the gauge symmetryies $U(4)_C$, $U(2)_L$ and $U(2)_R$.
 The gauge anomalies from three $U(1)$s are cancelled by
the generalized Green-Schwarz mechanism, and these $U(1)$s gauge fields
obtain masses via the linear $B\wedge F$ couplings. Thus, we obtain the
Pati-Salam gauge symmetries $SU(4)_C\times SU(2)_L\times SU(2)_R$.
Moreover, to have three families of the SM fermions,
 we require the intersection numbers to satisfy
\begin{eqnarray}
\label{E3LF} I_{ab} + I_{ab'}~=~3~,~\,
\end{eqnarray}
\begin{equation}
\label{E3RF} I_{ac} ~=~-3~,~ I_{ac'} ~=~0~,~\,
\end{equation}
where the conditions $I_{ab} + I_{ab'}=3$ and $I_{ac} =-3$ give us three generations of the SM fermions,
whose quantum numbers under $SU(4)_C\times SU(2)_L\times SU(2)_R$ gauge symmetries
are $({\bf 4, 2, 1})$ and $({\bf {\bar 4}, 1, 2})$. 
To satisfy the $I_{ac'} =0 $ condition, the stack $a$ D6-branes must be parallel to 
the orientifold ($\Omega R$) image $c'$ of the $c$-stack of D6-branes along at least one tow-torus,
where in our model building we choose to be the third two-torus. And then we have 
Open strings that stretch between the $a$ and $c'$ stacks of D6-branes. 
When the minimal distance square $Z^2_{(ac')}$ (in $1/M_s$ units) between these two stacks 
on the third two-torus is small, 
namely when the minimal length squared of the stretched string is small, we obtain the light scalars 
with squared-masses $Z^2_{(ab')}/(4\pi^2 \alpha')$ from the NS sector, and the light fermions 
with the same masses from R sector~\cite{Uranga,imr,LLG3},
which form four-dimensional $N=2$ hypermultiplets. Thus, we have $I_{ac'}^{(2)}$ 
(the intersection numbers for $a$ and $c'$ stacks on the first two two-tori) 
vector-like pairs of the chiral superfields with quantum numbers $({\bf {\bar 4}, 1, 2})$ and $({\bf 4, 1, 2})$. 
 These vector-like particles are the Higgs fields, which can break the
Pati-Salam gauge symmetry down to the SM gauge symmetry, while keep 
the four-dimensional $N=1$ supersymmetry. Especially, they are massless when $Z^2_{(ac')}=0$. 
Due to the symmetry transformation $c\leftrightarrow c'$,
 the model with intersection numbers $I_{ac}=0$ and $I_{ac'}=-3$ are
equivalent to that with $I_{ac}=-3$ and $I_{ac'}=0$, so we shall not discuss it here.

To break the Pati-Salam gauge symmetry to the SM, we split the $a$-stack of
D6-branes into $a_1$ and $a_2$ stacks respectively with 6 and 2 D6-branes.
And then the $U(4)_C$ gauge symmetry is broken down to $U(3)_C \times U(1)$. 
The gauge fields and three chiral multiplets in adjoint representation of $SU(4)_C$ are broken down to  
the gauge fields and three chiral multiplets in adjoint representations of $SU(3)_C$ and as well as the gauge field and three singlets of $U(1)_{B-L}$ accordingly. 
Also, we assume that the numbers of symmetric and anti-symmetric representations for $SU(4)_C$  
are $n_{\Ysymm}^a$ and $n_{\Yasymm}^a$, respectively, similar convention for $SU(3)_C$, $SU(2)_L$, and $SU(2)_R$.
These chiral multiplets for $SU(4)_C$ are broken down to the $n_{\Ysymm}^a$ and $n_{\Yasymm}^a$ chiral multiplets 
in symmetric and anti-symmetric representations for $SU(3)_C$, and $n_{\Ysymm}^a$ chiral multiplets 
with $U(1)_{B-L}$  charge $\pm 2$. Moreover, there exist $I_{a_1 a'_2}$ new fields 
with quantum number $({\bf 3, -1})$ under $SU(3)_C\times U(1)_{B-L}$ arising from the open strings 
at the intersections of $a_1$ and $a_2'$ stacks of D6-branes, while the rest of the particle spectrum 
remains the same. Also, the anomaly free gauge symmetries from $a_1$ and $a_2$ stacks of D6-branes 
are $SU(3)_C\times U(1)_{B-L}$, the $SU(4)_C$ subgroup.

To break $U(2)_R$ gauge symmetry,  we split the $c$-stack of D6-branes into
 $c_1$ and $c_2$ stacks, and each one has two D6-branes. And then the gauge fields and three chiral multiplets 
in adjoint representation of $SU(2)_R$ are broken down to the gauge field and three singlets of $U(1)_{I_{3R}}$, respectively. The $n_{\Ysymm}^c$ chiral multiplets 
in symmetric representation of $SU(2)_R$ are broken down to the $n_{\Ysymm}^c$ chiral multiplets 
with $U(1)_{I_{3R}}$ charge, while the $n_{\Yasymm}^c$ chiral multiplets 
in anti-symmetric representation $SU(2)_R$ will be gone. 
Also, there are $I_{c_1 c'_2}$ new fields that are neutral under $U(1)_{I_{3R}}$ arising from the open strings 
at the intersections of $c_1$ and $c_2'$ stacks of D6-brane, while the rest of the particle spectrum 
remain the same. The anomaly free gauge symmetry from $c_1$ and $c_2$ stacks of D6-branes 
becomes $U(1)_{I_{3R}}$, the $SU(2)_R$ Cartan subgroup.

With the above D6-brane splittings, we obtain the  
$SU(3)_C\times SU(2)_L\times U(1)_{B-L} \times U(1)_{I_{3R}}$ gauge symmetry. 
In order to break it further 
down to the SM gauge symmetry, we assume the minimal distance square $Z^2_{(a_2 c_1')}$ 
to be small, and thus obtain $I_{a_2 c_1'}^{(2)}$  pairs of chiral multiplets with quantum numbers 
$({\bf { 1}, 1, -1, 1/2})$ and $({\bf { 1}, 1, 1, -1/2})$ 
under $SU(3)_C\times SU(2)_L\times U(1)_{B-L} \times U(1)_{I_{3R}}$.  
These vector-like particles
 can break the $SU(3)_C\times SU(2)_L\times U(1)_{B-L} \times U(1)_{I_{3R}}$ gauge symmetry 
down to the SM while keep the D- and F-flatness since their quantum numbers are the same as 
those of the right-handed neutrino and its complex conjugate. In particular, they are massless 
when $Z^2_{(a_2c_1')}=0$.  Therefore, the complete chains for symmetry breaking are
\begin{eqnarray}
SU(4)\times SU(2)_L \times SU(2)_R  &&
\overrightarrow{\;a\rightarrow a_1+a_2\;}\;  SU(3)_C\times SU(2)_L
\times SU(2)_R \times U(1)_{B-L} \nonumber\\&&
 \overrightarrow{\; c\rightarrow c_1+c_2 \;} \; SU(3)_C\times SU(2)_L\times
U(1)_{I_{3R}}\times U(1)_{B-L} \nonumber\\&&
 \overrightarrow{\;\rm Higgs \;
Mechanism\;} \; SU(3)_C\times SU(2)_L\times U(1)_Y~.~\,
\end{eqnarray}
For Type IIA orientifolds with intersecting D6-branes, the dynamical supersymmetry breaking
has been studied in~Ref.~\cite{CLW}. There exist some filler branes carrying $USp$ gauge symmetries 
that are confining, and then could allow for gaugino condensation,
supersymmetry breaking, as well as moduli stabilization.

The gauge kinetic function for a generic stack $x$ of D6-branes is given by~\cite{CLW}
\begin{eqnarray}
f_x =  {\bf \textstyle{1\over 4}} \left[ n^1_x n^2_x n^3_x S -
(\sum_{i=1}^3 2^{-\beta_j-\beta_k}n^i_x l^j_x l^k_x U^i)  \right]
,\,
\end{eqnarray}
where the real parts of dilaton $S$ and moduli $U^i$ respectively are
\begin{eqnarray}
{\rm Re}(S) = \frac{M_s^3 R_1^{1} R_1^{2} R_1^{3} }{2\pi g_{s}}~,~\, \\
{\rm Re}(U^{i}) = {\rm Re}(S)~ \chi_j \chi_k~,~\,
\end{eqnarray}
where $i\neq j\neq k$,  and $g_s$ is the string coupling. 
So the gauge coupling constant associated with  $x$ is 
\begin{eqnarray}
\label{g2}
g_{D6_x}^{-2} &=& |\mathrm{Re}\,(f_x)|.
\end{eqnarray}
In our models, the holomorphic gauge kinetic functions for  $SU(4)_C$,  $SU(2)_L$
and $SU(2)_R$ are identified with stacks $a$,  $b$, and $c$, respectively.
The holomorphic gauge kinetic function for $U(1)_Y$ is then a linear combination of
 these for $SU(4)$ and $SU(2)_R$.
As shown in~\cite{bkl, Chen:2007zu}, we have
\begin{equation} \label{fy}
f_Y =  \frac{3}{5} \,( \frac{2}{3}\, f_{a} + f_{c} ).
\end{equation}
Also, we can express the tree-level MSSM gauge couplings in the form of
\begin{equation}\label{gy}
g^2_{a} = \alpha\, g^2_{b} = \beta\, \frac{5}{3}g^2_Y = \gamma\, \left[\pi e^{\phi_4}\right]
\end{equation}
where $g_a^2,  g^2_{b}$, and $\frac{5}{3}g^2_Y$ are the strong, weak and hypercharge gauge couplings, 
respectively, and $\alpha, \beta, \gamma$ are the ratios between them. 
Moreover, the K\"ahler potential is given by
\begin{eqnarray}
K=-{\rm ln}(S+ \bar S) - \sum_{I=1}^3 {\rm ln}(U^I +{\bar
U}^I).~\,
\end{eqnarray}
Three stacks of D6-branes, which carry $U(4)_C\times U(2)_L \times U(2)_R$ gauge symmetry,
 generically determine the complex structure moduli $\chi_1$, $\chi_2$ and $\chi_3$ 
because of the four-dimensional $N=1$ supersymmetry conditions. 
Thus,  we only have one independent modulus field. 
 In order to stabilize the moduli, 
one usually has at least two $USp$ groups with negative $\beta$ functions
which can be confined and then allow for gaugino condensations~\cite{Taylor,RBPJS,BDCCM}. 
In general, the one-loop beta function for the $2N^{(i)}$ filler branes, which are on top of $i$-th O6-plane 
and carry $USp(N^{(i)})$ group, is given by~\cite{Cvetic:2004ui}
\begin{eqnarray}
\beta_i^g&=&-3({N^{(i)}\over2}+1)+2 |I_{ai}|+
|I_{bi}| +  |I_{ci}|
+3({N^{(i)}\over2}-1)\nonumber\\
        &=&-6+2 |I_{ai}|+  |I_{bi}|+  |I_{ci}|~.~\,
\label{betafun}
\end{eqnarray}
If supersymmetry is broken by gaugino condensations, 
we may need to consider gauge mediation since gravity mediation is much smaller. 
Thus, the supersymmetry CP problem may be solved as well. 
 Unlike Ref.~\cite{Cvetic:2004ui} to include alternative supersymmetry broken mechanisms, we will not require at least two $USp$ gauge group factors with negative $\beta$ functions in our Pati-Salam model building.

From the pheonomenological point of view, we want to emphasize that 
Type II T-duality in Ref.~\cite{Cvetic:2004ui}
is not an equivalent relation in Pati-Salam model building 
if the model is not invariant under $SU(2)_L$ and $SU(2)_R$ exchange. 
Under the Type II T-duality,  the transformations of the wrapping
numbers for any stacks of D6-branes in the model are
\begin{eqnarray}
n_x^i \rightarrow -n_x^i,~ l_x^i \rightarrow l_x^i,&
 n_x^j \leftrightarrow l_x^j,&
n_x^k \leftrightarrow l_x^k,\, \label{T-duality II}
\end{eqnarray}
where $i\not= j \not= k$, as well as 
 $x$ runs over all D6-branes in the model.
In particular, it is easy to show that all the intersection numbers will change signs.

For a three-family supersymmetric Pati-Salam model,
 we obtain a corresponding new three-family supersymmetric Pati-Salam models
by exchanging $b$-stacks and $c$-stacks of D6-branes
\begin{eqnarray}
b \leftrightarrow c~.~\,
\end{eqnarray}
Especially, the quantum numbers for
$SU(2)_L$ and $SU(2)_R$ in the particle spectrum, as well as the $SU(2)_L$ and $SU(2)_R$ gauge
couplings at string scale will be 
interchanged due to the $b$-stacks and $c$-stacks exchange
$b \leftrightarrow c$. 
Therefore, from the phenomenological point of view, this is not an equivalent relation
if the particle content is not invariant under $SU(2)_L$ and $SU(2)_R$ exchange 
or if $SU(2)_L\times SU(2)_R$ gauge couplings are not unified at the string scale.
To be more concise, this is not an equivalent relation if the Pati-Salam model is not invariant 
under $SU(2)_L$ and $SU(2)_R$ exchange.

Moreover, there is a variation of type II T-duality~\cite{Cvetic:2004ui}. 
Under it, the transformations of
the wrapping numbers for any stacks of D6-branes in the model are
\begin{eqnarray}
l_x^1 \rightarrow -l_x^1,& l_x^2 \rightarrow -l_x^2,& l_x^3
\rightarrow -l_x^3, \nonumber\\
b\leftrightarrow c, \label{T-duality IIa}
\end{eqnarray}
where $x$ runs over all D6-branes in the model. 

This leads to one interesting phenomenon on gauge coupling unification aspect.
Looking closer at the gauge kinetic function relation Eq.~\eqref{fy} and the 
MSSM gauge coupling relation Eq.~\eqref{gy}, it is obvious that when $f_a/f_b =1$, 
namely when there is the gauge coupling unification for the $SU(4)_C$  and $SU(2)_L$ 
gauge symmetries,  after the $b$- and $c$-stacks of brane swapped,  it will be shifted to 
the $U(1)_Y $ and $SU(4)_C$ gauge coupling unification, {\it i.e.}, $f_Y/f_a =1$. 
Similarly, when there is a $U(1)_Y $ and $SU(4)_C$ gauge coupling unification 
before the $b$- and $c$-stacks of brane swapping, it will be shifted to 
$SU(4)_C$ and $SU(2)_L$ gauge coupling unification at string scale.

\subsection{Scanning of Supersymmetric Pati-Salam Models}

We shall search for the new Pati-Salam models with basic properties in the previous subsection. 
Similar to Ref.~\cite{Cvetic:2004ui}, we introduce three stacks of D6-branes, $a$, $b$, and $c$ 
with number of D6-branes 8, 4, and 4, respectively. The corrsponding gauge symmetries are
 $U(4)_C$, $U(2)_L$ and $U(2)_R$. 
Unlike the strategy in Ref.~\cite{Cvetic:2004ui}, we do not restrict ourselves with 
that at least two $USp$ groups in the hidden sector have negative $\beta$ functions, 
and instead we do a broader scanning without any constraint on the hidden sector.

In general, if all three two-tori are not tilted, we can not obtain the particle
spectra with odd generations of the SM fermions. Thus, we have three
kinds of scenarios: one tilted two-torus, two tilted two-tori, and
three tilted two-tori.
As pointed out in Ref.~\cite{Cvetic:2004ui},  the model buildings 
with two and three tilted two-tori either do not have
 three families or violate the RR tadpole cancellation conditions. And our scanning confirms
this observation. Therefore, here we concentrate on the new scanning with only one tilted torus. 
We choose the third two-torus to be tilted and study the new inequivalent Pati-Salam models in the following.
In our broader scanning, we obtain several classes of new models. 

The first class of models including Models \ref{oneOplane1}, \ref{oneOplane01}, \ref{oneOplane2}, 
and \ref{oneOplane02} has only one $USp$ group,  in which Models \ref{oneOplane1} and \ref{oneOplane01}, 
as well as Models \ref{oneOplane2} and \ref{oneOplane02} are T-dual to each other. 
Especially, the Models \ref{oneOplane2} and \ref{oneOplane02} do not have the colored chiral exotic particles.
The Higgs particles in Models \ref{oneOplane1} and \ref{oneOplane2} 
arise from $N=2$ subsectors
at the intersections of $b$- and $c$-stacks of D6-branes, 
while the Higgs particles in Models \ref{oneOplane01} and \ref{oneOplane02}
arise from $N=2$ subsectors
at the intersections of $b$- and $c'$-stacks of D6-branes.
There exist four and eight exotic Higgs-like particles in 
Models \ref{oneOplane1} and \ref{oneOplane01} as well as 
 Models \ref{oneOplane2} and \ref{oneOplane02}, respectively.

The second class of models has two $USp$ groups, and the representative models are
Models \ref{twoOplane1} and  \ref{twoOplane01},
which are T-dual to each other. The Higgs particles in Model \ref{twoOplane1}
arise from $N=2$ subsector
at the intersections of $b$- and $c$-stacks of D6-branes, 
while the Higgs particles in Models \ref{twoOplane01} 
arise from $N=2$ subsectors
at the intersections of $b$- and $c'$-stacks of D6-branes.
Because all these $USp$ groups have negative $\beta$ functions, 
we may stabilize the modulus and break the supersymmetry via gaugino condensations.

The third class of models has more than two $USp$ groups, and the representative models are
Models  \ref{threeOplane1}, \ref{threeOplane01}, \ref{fourOplane1}, and \ref{fourOplaneT}.
There exist at least two $USp$ groups in these models, which have negative 
$\beta$ functions. So we may stabilize the modulus and break the supersymmetry 
via gaugino condensations as well. 
Model \ref{fourOplaneT}, which have four confining $USp(N)$ gauge groups 
and can considered as T-dual of Model  I-Z-10 in \cite{Cvetic:2004ui}, 
is the only model in our current scan which has exact gauge coupling unification at the string scale.

The fourth class of models has the absolute value of at least one wrapping number equal to $5$, 
and the representative models are Models \ref{5wrap1} and \ref{5wrap2}.
This kind of models has not been found in the previous search~\cite{Cvetic:2004ui}.
Interestingly,  we observe that the MSSM gauge coupling values are more refined, and
there exists the approximate gauge coupling unification.
Also,  Model \ref{5wrap2} has two $USp$ groups with negative $\beta$ functions. 
While in Model \ref{5wrap1} there is only one $USp$ group with negative $\beta$ function,
so we might need to stabilize the modulus with different mechanism. 

The fifth class of models has at least one wrapping number whose absolute value is larger than or equal to $5$, 
and the representative models are Models \ref{model5} - \ref{model2}. We revise our random 
scanning methods with supervised scanning and raise the scanning efficiency. A similar cluster 
behaviours of three-family models can also be observed in the later discussions in Section~\ref{sec:plot} 
as for the models with small wrapping numbers. In the next Section we will focus on the 
phenomenological studies of the models with small wrapping numbers.

\section{Preliminary Phenomenological Studies}
\label{sec:pheno}

In this section, we shall discuss the phenomenological features of our models. 
We start with Models \ref{oneOplane1} and \ref{oneOplane01}, 
which are constructed with one $USp$ group. The gauge symmetry is
$U(4)\times U(2)_L\times U(2)_R\times USp(4)$, while the $\beta$ function of $USp(4)$ group
is zero. So we cannot break supersymmetry via gaugino condensation, and then need the
other mechanism for supersymmetry breaking. 
Also, how to decouple the exotic particles, which are charged under $USp(4)$, is 
an interesting question since $USp(4)$ is not confined. 
In Models \ref{oneOplane2} and \ref{oneOplane02}, the gauge symmetry is
$U(4)\times U(2)_L\times U(2)_R\times USp(4)$ as well.
The $\beta$ function of $USp(4)$ group is negative, so we can break supersymmety
via gaugino condensation, and decouple the exotic particles.
In all these models, we need to address the modulus stabilization issue as well,
which is generic for the models with one $USp$ group.

\begin{table}
	[htb] \footnotesize
	\renewcommand{\arraystretch}{1.0}
	\caption{The chiral spectrum in the open string sector for Model \ref{oneOplane1}. } 
	\label{oneplane1}
	\begin{center}
		\begin{tabular}{|c||c||c|c|c||c|c|c|}\hline
			\ref{oneOplane1} & $SU(4)\times SU(2)_L\times SU(2)_R \times USp(4)$
			& $Q_4$ & $Q_{2L}$ & $Q_{2R}$ & $Q_{em}$ & $B-L$ & Field \\
			\hline\hline
			$ab$ & $3 \times (4,\overline{2},1,1)$ & 1 & $-1$ & 0  & $-\frac 13,\; \frac 23,\;-1,\; 0$ & $\frac 13,\;-1$ & $Q_L, L_L$\\
			$ac$ & $3 \times (\overline{4},1,2,1)$ & $-1$ & 0 & $1$   & $\frac 13,\; -\frac 23,\;1,\; 0$ & $-\frac 13,\;1$ & $Q_R, L_R$\\
						$bc'$ & $4 \times(1,\overline{2}, \overline{2},1)$ & 0 & $-1$ & $-1$   & $-1,\;0,\;0,\;1$ & 0 & $H'$\\
$a2$ & $2\times (\overline{4},1,1,4)$ & $-1$ & 0 & 0   & $-\frac 16,\;\frac 12$ & $-\frac 13,\;1$ & \\
			$b2$ & $1\times(1,2,1,\overline{4})$ & 0 & 1 & 0   & $\pm \frac 12$ & 0 & \\
$c2$ & $1\times(1,1,\overline{2},4)$ & 0 & 0 & -1  & $\mp \frac 12$ & 0 & \\
			$a_{\Ysymm}$ & $1\times(10,1,1,1)$ & 2 & 0 & 0   & $\frac 13, -\frac 13,-1$ & $\frac 23,-2$ & \\
$a_{\Yasymm}$ & $1\times(\overline{6},1,1,1)$ & -2 & 0 & 0   & $-\frac 13, 1$ & $-\frac 23,2$ & \\
			$b_{\Ysymm}$ & $2\times(1,3,1,1)$ & 0 & $2$ & 0   & $0,\pm 1$ & 0 & \\
			$b_{\overline{\Yasymm}}$ & $2\times(1,{1},1,1)$ & 0 & -2 & 0   & 0 & 0 & \\
			$c_{\overline{\Ysymm}}$ & $2\times(1,1,\overline{3},1)$ & 0 & 0 & -2   & $0,\pm 1$ & 0 & \\
			$c_{\Yasymm}$ & $6\times(1,1,{1},1)$ & 0 & 0 & -2   & 0 & 0 & \\
	\hline\hline
$bc$ & $8 \times (1,2,\overline{2},1)$ & 0 & 1 & -1   & $1,\;0,\; 0,\;-1$ &0 &$H_u^i, H_d^i$\\
& $8 \times (1,\overline{2},2,1)$ & 0 & -1 & 1   &  & &  \\
	\hline
		\end{tabular}
	\end{center}
\end{table}

\begin{table}
	[htb] \footnotesize
	\renewcommand{\arraystretch}{1.0}
	\caption{The chiral spectrum in the open string sector for Model \ref{oneOplane01}. } 
		\label{oneplane01}
	\begin{center}
		\begin{tabular}{|c||c||c|c|c||c|c|c|}\hline
			\ref{oneOplane01} & $SU(4)\times SU(2)_L\times SU(2)_R \times USp(4)$
			& $Q_4$ & $Q_{2L}$ & $Q_{2R}$ & $Q_{em}$ & $B-L$ & Field \\
			\hline\hline
		$ab'$ & $3 \times (4,2,1,1)$ & 1 & 1 & 0  & $-\frac 13,\; \frac 23,\;-1,\; 0$ & $\frac 13,\;-1$ & $Q_L, L_L$\\
			$ac$ & $3 \times (\overline{4},1,2,1)$ & -1 & 0 & 1   & $\frac 13,\; -\frac 23,\;1,\; 0$ & $-\frac 13,\;1$ & $Q_R, L_R$\\
	 $bc$ & $4 \times(1,2,\overline{2},1)$ & 0 & 1 & -1   & $1,\;0,\;0,\;-1$ & 0 & $H$\\
			$a2$ & $2\times (\overline{4},1,1,4)$ & -1 & 0 & 0   & $-\frac 16,\;\frac 12$ & $-\frac 13,\;1$ & \\
			$b2$ & $1\times(1,\overline{2},1,4)$ & 0 & -1 & 0   & $\mp \frac 12$ & 0 & \\
			$c2$ & $1\times(1,1,\overline{2},4)$ & 0 & 0 & -1  & $\mp \frac 12$ & 0 & \\
			$a_{\Ysymm}$ & $1\times(10,1,1,1)$ & 2 & 0 & 0   & $\frac 13, -\frac 13,-1$ & $\frac 23,-2$ & \\
$a_{\Yasymm}$ & $1\times(\overline{6},1,1,1)$ & -2 & 0 & 0   & $-\frac 13, 1$ & $-\frac 23,2$ & \\
			$b_{\Ysymm}$ & $2\times(1,\overline{3},1,1)$ & 0 & -2 & 0   & $0,\pm 1$ & 0 & \\
			$b_{\overline{\Yasymm}}$ & $2\times(1,{1},1,1)$ & 0 & 2 & 0   & 0 & 0 & \\
			$c_{\overline{\Ysymm}}$ & $2\times(1,1,\overline{3},1)$ & 0 & 0 & -2   & $0,\pm 1$ & 0 & \\
			$c_{\Yasymm}$ & $6\times(1,1,{1},1)$ & 0 & 0 & -2   & 0 & 0 & \\
			\hline
		\end{tabular}
	\end{center}
\end{table}

The Models \ref{twoOplane1} and \ref{twoOplane01}  are with two confining gauge groups 
in hidden sector. They are T-dual to each other 
in the same manner as for the one orientifold plane class of models, and they can be obtained  
from Model I-Z-2 in Ref.~\cite{Cvetic:2004ui} via generic Type II duality. 
The full spectrum of Model \ref{twoOplane01} is shown in Table \ref{twooplane}. 
Because we do not have $SU(2)_L \times SU(2)_R$ gauge coupling unification at the string scale,
the Models \ref{twoOplane1} and \ref{twoOplane01} are not equivalent to the Model I-Z-2 in Ref.~\cite{Cvetic:2004ui}
from phenomenological point of view. At the string scale,
we have $SU(4)_C\times SU(2)_R$ gauge coupling unification in Models \ref{twoOplane1} and \ref{twoOplane01},
while  $SU(4)_C\times SU(2)_L$ gauge coupling unification in Model I-Z-2 of \cite{Cvetic:2004ui}.
Recalling the definition of hypercharge gauge coupling, we find that 
the $SU(3)_C \times SU(2)_L$ gauge coupling unification is shifted to $SU(3)_C \times U(1)_Y$ gauge coupling unification,
and vice versa.
Similarly, the Model \ref{fourOplane1} has $SU(3)_C \times SU(2)_L$ gauge coupling unification
 while its corresponding model, which is constructed via generic Type II duality, has 
$SU(3)_C \times U(1)_Y$ gauge coupling unification.
Thus, the generic Type II duality provides a new way to construct the new models.
For Model \ref{threeOplane1} with approximate $SU(3)_C \times U(1)_Y$ gauge coupling unification,  
we obtain Model \ref{threeOplane01} with approximate $SU(3)_C \times SU(2)_L$ gauge coupling unification
via generic Type II T-duality. 
However, we should note that this construction is not simply swapping the $b$- and $c$-stack of D6-branes, 
but usually the non-trivial Type II T-dualities are performed. 
 For the examples of performing  $b$- and $c$-stack of D6-branes swapping under Eq.~\eqref{T-duality IIa}, 
we show Models \ref{5wrap01}, \ref{5wrap11}, and \ref{5wrap21} respectively from 
Models \ref{5wrap0}, \ref{5wrap1}, and \ref{5wrap2} via Type II duality, 
in which the SM gauge couplings are shifted resulting from such D6-brane swapping. 

\begin{table}
	[htb] \footnotesize
	\renewcommand{\arraystretch}{1.0}
	\caption{The chiral spectrum in the open string sector for Model \ref{twoOplane01}.} \label{twooplane}
	\begin{center}
		\begin{tabular}{|c||c||c|c|c||c|c|c|}\hline
			\ref{twoOplane01} & $SU(4)\times SU(2)_L\times SU(2)_R \times 	USp(4)^2 $
			& $Q_4$ & $Q_{2L}$ & $Q_{2R}$ & $Q_{em}$ & $B-L$ & Field \\
			\hline\hline
			$ab$ & $3 \times (4,\overline{2},1,1,1)$ & 1 & -1 & 0  & $-\frac 13,\; \frac 23,\;-1,\; 0$ & $\frac 13,\;-1$ & $Q_L, L_L$\\
	$ac'$ & $3 \times (\overline{4},1,\overline{2},1,1)$ & -1 & 0 & -1   & $\frac 13,\; -\frac 23,\;1,\; 0$ & $-\frac 13,\;1$ & $Q_R, L_R$\\
	$bc$ & $4 \times(1,2,\overline{2},1,1)$ & 0 & 1 & -1   & $-1,\;0,\;0,\;1$ & 0 & $H$\\
			$a3$ & $1\times (\overline{4},1,1,4,1)$ & -1 & 0 & 0   & $-\frac 16,\;\frac 12$ & $-\frac 13,\;1$ & \\
		$a4$ & $1\times (4,1,1,1,\overline{4})$ & $1$ & 0 & 0   & $\frac 16,\;-\frac 12$ & $\frac 13,\;-1$ & \\
			$b3$ & $2\times (1,2,1,\overline{4},1)$ & $0$ & 1 & 0   & $\pm\frac 12$ & 0 & \\
			$b4$ & $1\times(1,\overline{2},1,1,4)$ & 0 & -1 & 0   & $\mp \frac 12$ & 0 & \\
		$c4$ & $1\times(1,1,2,1,\overline{4})$ & 0 & 0 & 1   & $\pm \frac 12$ & 0 & \\
			$b_{\Ysymm}$ & $2\times(1,\overline{3},1,1,1)$ & 0 & -2 & 0   & $0,\pm 1$ & 0 & \\
			$b_{\overline{\Yasymm}}$ & $6\times(1,{1},1,1,1)$ & 0 & -2 & 0   & 0 & 0 & \\
			$c_{\overline{\Ysymm}}$ & $2\times(1,1,{3},1,1)$ & 0 & 0 & 2   & $0,\pm 1$ & 0 & \\
			$c_{\Yasymm}$ & $2\times(1,1,{1},1,1)$ & 0 & 0 & -2   & 0 & 0 & \\
			\hline
		\end{tabular}
	\end{center}
\end{table}

\begin{table}
	[htb] \footnotesize
	\renewcommand{\arraystretch}{1.0}
	\caption{The chiral spectrum in the open string sector for Model  \ref{5wrap2}.} \label{spectrum5wrap2}
	\begin{center}
		\begin{tabular}{|c||c||c|c|c||c|c|c|}\hline
			 \ref{5wrap2} & $SU(4)\times SU(2)_L\times SU(2)_R  \times	USp(2)^3$
			& $Q_4$ & $Q_{2L}$ & $Q_{2R}$ & $Q_{em}$ & $B-L$ & Field \\
			\hline\hline
	$ab'$ & $3 \times (4,2,1,1,1,1)$ & 1 & 1 & 0  & $-\frac 13,\; \frac 23,\;-1,\; 0$ & $\frac 13,\;-1$ & $Q_L, L_L$\\
			$ac$ & $3 \times (\overline{4},1,2,1,1,1)$ & $-1$ & 0 & $1$   & $\frac 13,\; -\frac 23,\;1,\; 0$ & $-\frac 13,\;1$ & $Q_R, L_R$\\
		$bc$ & $1 \times(1,\overline{2},2,1,1,1)$ & 0 & $-1$ & $1$   & $1,\;0,\;0,\;-1$ & 0 & $H$\\
			$a2$ & $1\times (\overline{4},1,1,2,1,1)$ & $-1$ & 0 & 0   & $-\frac 16,\;\frac 12$ & $-\frac 13,\;1$ & \\
		$a4$ & $1\times (4,1,1,1,1,\overline{2})$ & $1$ & 0 & 0   & $\frac 16,\;-\frac 12$ & $\frac 13,\;-1$ & \\
			$b2$ & $1\times(1,\overline{2},1,2,1,1)$ & 0 & -1 & 0   & $\mp \frac 12$ & 0 & \\
				$b3$ & $2\times (1,2,1,1,\overline{2},1)$ & $0$ & 1 & 0   & $\pm\frac 12$ & 0 & \\
					$b4$ & $2\times(1,2,1,1,1,\overline{2})$ & 0 & 1 & 0   & $\pm \frac 12$ & 0 & \\
			$c3$ & $5\times(1,1,2,1,\overline{2},1)$ & 0 & 0 & 1   & $\pm \frac 12$ & 0 & \\
			$b_{\overline{\Yasymm}}$ & $8\times(1,{1},1,1,1,1)$ & 0 & 2 & 0   & 0 & 0 & \\
			$c_{\overline{\Ysymm}}$ & $3\times(1,1,\overline{3},1,1,1)$ & 0 & 0 & -2   & $0,\pm 1$ & 0 & \\
			$c_{\Yasymm}$ & $3\times(1,1,1,1,1,1)$ & 0 & 0 & 2   & 0 & 0 & \\
			\hline
		\end{tabular}
	\end{center}
\end{table}

As we mentioned, the $USp(N)$ gauge groups with negative beta functions in hidden sector have a potential to be confining, 
and thus  the non-perturbative effective superpotential can be generated 
via gaugino condensations.  The ground state, which is determined by the minimization of this supergravity potential, 
can stabilize the dilaton and complex structure toroidal moduli, and breaks supersymmetry in some cases.  
For the models with two confining $USp(N)$ gauge groups, a general analysis of 
the non-perturbative superpotential with tree-level gauge couplings can be performed,
and it was shown that there can exist extrema with the stabilizations of dilaton and complex structure moduli~\cite{CLW}. 
 However, these extrema might be saddle points and thus do not break supersymmetry. Interestingly,
if the models have three or four confining $USp(N)$ gauge groups,  the non-perturbative superpotientil 
allows for the  moduli stabilization and supersymmetry breaking at the stable extremum in general~\cite{CLW}. 

Among our representative models, two Models (\ref{oneOplane1} and \ref{oneOplane01}) carry 
one  $USp(N)$ gauge group with zero beta function, three Models (\ref{oneOplane2}, \ref{oneOplane02}, 
and \ref{5wrap1}) have one confining $USp(N)$ gauge group with negative beta function, 
six Models (\ref{twoOplane1}, \ref{twoOplane01}, \ref{threeOplane1}, \ref{threeOplane01}, \ref{fourOplane1}, 
and \ref{5wrap2}) carry two confining $USp(N)$ gauge groups with negative beta functions, 
and one Model \ref{fourOplaneT} have four confining $USp(N)$ gauge groups and considered as T-dual of Model  I-Z-10 in \cite{Cvetic:2004ui}. Therefore, for the latter seven models, there may exist the
 stable extrema with  moduli stabilization and supersymmetry breaking due to gaugino condensations,
which are very interesting from the phenomenological points of view. 
However, as pointed out in Ref.~\cite{CLW}, the cosmological constants at these extrema are likely 
to be negative and close to the string scale, and thus 
the gaugino condensations in these models might not address the cosmological constant problem.

All the models contain the exotic particles that are charged under the hidden gauge groups. 
The strong coupling dynamics in hidden sector at certain intermediate scale 
might provide a mechanism for all these particles to form bound states or composite particles, 
which are compatible with anomaly cancellation conditions. 
And then similar to the quark condensation in QCD,
these particles would be only charged under the SM gauge symmetry~\cite{CLS1}.
The $USp$ groups have two kinds of neutral bound states in general. 
The first one is the pseudo inner product of two fundamental representations that is generated by decomposing
the rank two anti-symmetric representation, and is general for $USp$ groups. In some sense, this is 
the reminiscent of a meson that is the inner product of one pair of fundamental and 
anti-fundamental representations of $SU(3)_C$ in QCD. 
The second one is the rank $2N$ anti-symmetric representation of $USp(2N)$ group for $N\geq 2$, 
which is an $USp(2N)$ singlet and somewhat similar to a  baryon, 
as a rank three anti-symmetric representation of $SU(3)_C$  in QCD. 
Our models, which contain the second kind of neutral bound states, 
are Models \ref{oneOplane2} and \ref{oneOplane02} with confining $USp(4)$ group, 
as well as Models \ref{twoOplane1} and \ref{twoOplane01} with confining groups $USp(4)\times USp(4)$ in the hidden sector.  
We note that Model \ref{twoOplane1} and \ref{twoOplane01} are T-dual to each other,
 and are constructed by their $b$- and $c$-stack of D6-branes swapped 
from Model I-Z-2 in \cite{Cvetic:2004ui} with proper T-duality transformations.  
For $N=1$, these two kinds are the same. 

Now we take Models \ref{oneOplane02} and \ref{twoOplane01} as examples to show explicitly 
the new composite states. In Model \ref{oneOplane02}, we present the confined particle
spectrum in Table \ref{Composite Particles oneOplane02}. Because it has 
one confining gauge group $USp(4)$ with two charged intersections. Therefore, 
besides self-confinement, the mixed-confinement between different intersections 
is also possible, which yields the chiral supermultiplets $(1,2, \overline{2},1)$.
In Model \ref{twoOplane01}, the confined particle spectra are given 
in Table \ref{Composite Particles twoOplane01}. It has two confining gauge groups $USp(4)_3$ and $USp(4)_4$ 
both with two charged intersections. Besides the self-confinement, the mixed-confinement 
between different intersections yields the chiral supermultiplets 
$(\overline{4},2,1,1,1), (4,\overline{2},1,1,1), (1,\overline{2},2,1,1)$, and $(4,1,2,1,1)$.
Note that when there is only one charged intersection, 
we do not have mixed-confinement, and only the tensor representations are yielded from self-confinement. 
Moreover, it is easy to check from the spectrum that no new anomaly is introduced to 
the remaining gauge symmetry, so this model is still anomaly free. 

\begin{table}
	[htb] \footnotesize
	\renewcommand{\arraystretch}{1.0}
	\caption{The composite particle spectrum for Model \ref{oneOplane02}
		formed due to the strong forces in hidden sector.}
	\label{Composite Particles oneOplane02}
	\begin{center}
		\begin{tabular}{|c|c||c|c|}\hline
			\multicolumn{2}{|c||}{Model \ref{oneOplane02}} &
			\multicolumn{2}{c|}{$SU(4)\times SU(2)_L\times SU(2)_R \times
				USp(4)$} \\
			\hline Confining Force & Intersection & Exotic Particle
			Spectrum & Confined Particle Spectrum \\
			\hline\hline
			$USp(4)_2$ &$b2$ & $1\times(1,2,1,\overline{4})$ & $1\times(1,1,1,1)$, $1\times(1,3,1,1)$, $3\times(1,2, \overline{2},1)$\\
			&$c2$ & $3\times(1,1,\overline{2},4)$ & $6\times(1,1,1,1)$, $6\times(1,1,\overline{3},1)$\\
			\hline
		\end{tabular}
	\end{center}
\end{table}

\begin{table}
	[htb] \footnotesize
	\renewcommand{\arraystretch}{1.0}
	\caption{The composite particle spectrum for Model \ref{twoOplane01}
		formed due to the strong forces in hidden sector.}
	\label{Composite Particles twoOplane01}
	\begin{center}
		\begin{tabular}{|c|c||c|c|}\hline
			\multicolumn{2}{|c||}{Model \ref{twoOplane01}} &
			\multicolumn{2}{c|}{$SU(4)\times SU(2)_L\times SU(2)_R \times
				USp(4)^2$} \\
			\hline Confining Force & Intersection & Exotic Particle
			Spectrum & Confined Particle Spectrum \\
			\hline\hline
			$USp(4)_3$ &$a3$ & $1\times  (\overline{4},1,1,4,1)$ & $1\times  (\overline{6},1,1,1,1)$,$1\times  (\overline{10},1,1,1,1)$, $2\times(\overline{4},2,1,1,1)$\\
			&$b3$ & $2\times(1,2,1,\overline{4},1)$ & $4\times(1,1,1,1,1)$, $4\times(1,3,1,1,1)$\\
			\hline
			$USp(4)_4$ &$a4$ & $1\times (4,1,1,1,\overline{4})$ & $1\times (6,1,1,1,1)$, $1\times (10,1,1,1,1)$, $1\times(4,\overline{2},1,1,1)$\\
			&$b4$ & $1\times(1,\overline{2},1,1,4)$ &  $1\times(1,{1},1,1,1)$,$1\times(1,\overline{3},1,1,1)$, $1\times(1,\overline{2},2,1,1)$\\
			&$c4$ & $1\times(1,1,2,1,\overline{4})$ &  $1\times(1,1,1,1,1)$, $1\times(1,1,3,1,1)$, $1\times(4,1,2,1,1)$\\
			\hline
		\end{tabular}
	\end{center}
\end{table}

This kind of self-confinement and mixed-confinement between different intersections also applies to 
the other models except for the Models \ref{oneOplane1} and \ref{oneOplane01}.
In these two models, we do not have asymptotical free gauge symmetries in the hidden sector, 
so the states charged under these symmetries cannot be confined. 
Because the anomaly cancellations for the confined particle spectra are not automatically guaranteed,
 one extra field associated with composite states may be needed to satisfy t'~Hooft anomaly matching condition. 
To avoid the unnecessary complications, we only consider 
relatively simple examples here.

\section{Machine Learning and Future Model Building}
\label{sec:plot}

In this section, we briefly review our scanning methods and employing dimension reduction methods visually show the hints for future model building. 
Firstly, we employed the standard random scanning according to the wrapping numbers up to $5$. This is efficient for the scanning with small wrapping numbers less than $5$, while it became very low efficient from wrapping number $5$. We improved our scanning methods by setting at least one wrapping number to be large\,(larger than 5), and setting the rest scanning to be random. In such a supervised way, we choose for one torus of one stack of brane\,(a, b or c-stack of brane) is constructed with larger wrapping number, while the other torus are constructed with random wrapping~(normally with small wrapping but wrapping number $5$ or more also appear) to have a MSSM-like physics model as we discussed in Section~\ref{sec:models}. 
Based on the above models, we expect it would be interesting to perform machine learning methods by feeding the collected data of the above models\,(e.g. the wrapping numbers and the intersection numbers) from supervised and random scanning to neural network, etc. In this way, one can train the neural network to select the MSSM-like physics models from the random constructed models and generate more MSSM-like physics models. 

\begin{figure}[h!]
	\centering
	\includegraphics[scale=1.2]{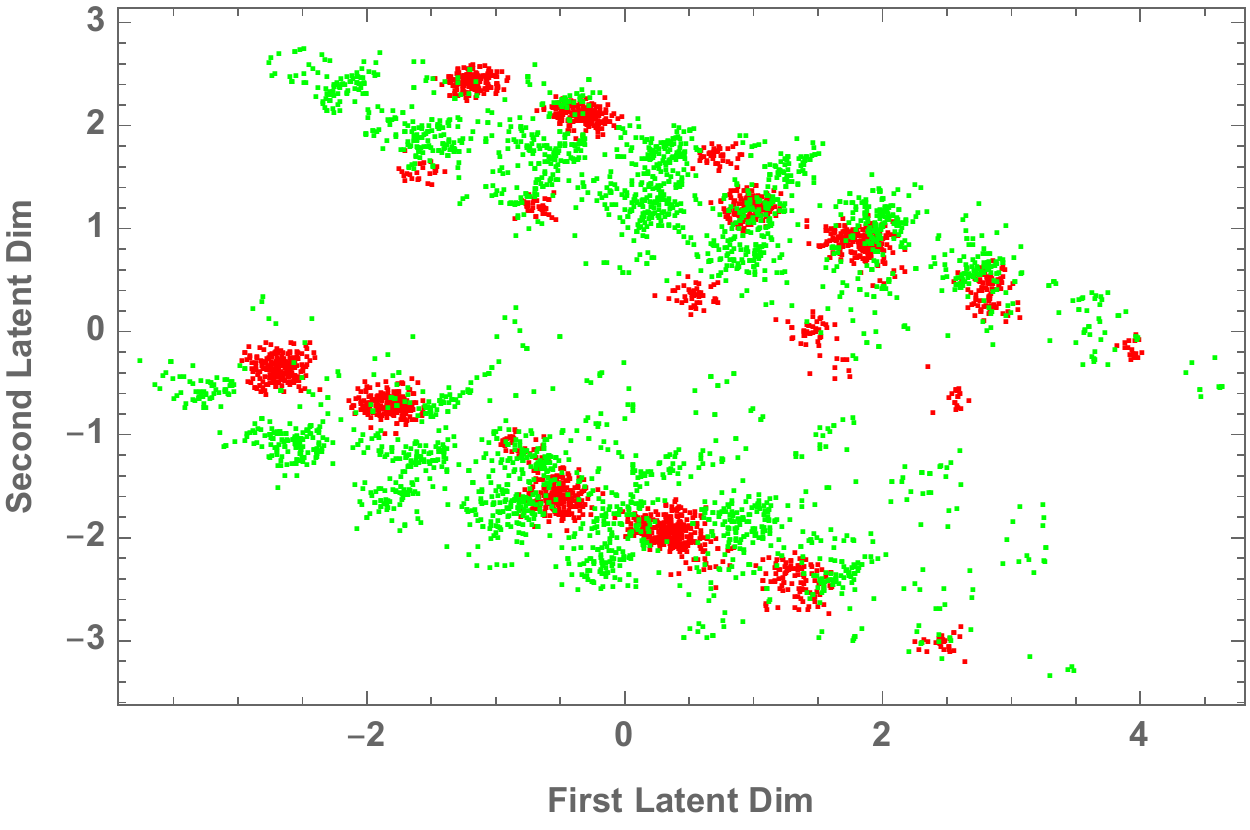}
	\caption{\label{fig:landscape1}LatentSemanticAnalysis Dimension Reduction}
\end{figure}

Now to discuss the possible improvement for future model building, we employ the dimension reduction methods and visually show the expansion of our constructed models. 
Firstly,  we visually show how the constructed models with three families of the SM fermions expand with dimension reduction method ``LatentSemanticAnalysis''.
With the so-called dimension reduction methods, we reduce the total $18$ wrapping numbers with $18$-dimension to $2$-dimension and observe that the selected MSSM-like models expand with pattern according to the first and second latent dimensions as shown in Figure~\ref{fig:landscape1}. 
In which, each point corresponds to a D6-brane model, and the MSSM-like models with their wrapping number less than 5 are highlighted with green, 
while the MSSM-like models with the wrapping number is larger than or equal to 5 are highlighted with red.
It is clear that the MSSM-like models trend to gather on some islands. Recall that the dimensions are reduced from the wrapping number, we expect once some MSSM-like models are obtained, scanning around the obtained MSSM-like models will increase the efficiency for future model building.  

In an alternative way, with dimension reduction method ``AutoEncoder'', we also observe that  according to the wrapping numbers of D6-brane models: each point corresponds to a D6-brane model, and the MSSM-like models are highlighted with green points when the wrapping number is less than 5, 
while the MSSM-like models are highlighted with red points when the wrapping number is larger than or equal to 5. A similar observation that the MSSM-like models populate in several separated islands also appear. 
The fact, where the MSSM-like models gather on islands in both methods, indicates that there will be more chances to construct new MSSM-like models there. We note that the dense observation  partially contains the contributions from the T-dual MSSM-like models constructed in the nearby region. With linear algorithm, this behaviors, which the MSSM-like models tend to gather, can be confirmed as well. 

Furthermore, we would like to mention that the heterotic orbifold landscape drawn with deep autoencoder neural network methods has similar cluster behaviour as shown in~\cite{MPV}. It seems to us that there might be such cluster behaviors or tend to gather for different model buildings. And thus, perturb around the current scanning might further enrich the obtained MSSM-like models in different model buildings.

\section{Discussions and Conclusions}
\label{sec:conclusion}

We  revisited  the systematic construction of the three-family $N=1$
supersymmetric Pati-Salam models from Type IIA orientifolds on
$\IT^6/(\IZ_2\times \IZ_2)$ with intersecting D6-branes,
where the $SU(4)_C\times SU(2)_L \times SU(2)_R$ gauge symmetries 
arise from the stacks of D6-branes with  $U(n)$ gauge symmetries.
We found that the Type II T-duality in Ref.~\cite{Cvetic:2004ui} is not
an equivalent relation in Pati-Salam model building if the model is not invariant under $SU(2)_L$ 
and $SU(2)_R$ exchange, and provides a way to obtain the new model.
Unlike the previous studies, we did not require at least two confining $USp$ groups. 
Also, we scanned the wrapping numbers up to $5$, and 
obtained more interesting models with approximate gauge coupling unification.
The Pati-Salam gauge symmetry can be broken down to the 
$SU(3)_C\times SU(2)_L\times U(1)_{B-L} \times U(1)_{I_{3R}}$ via D6-brane splittings, 
and further down to the SM gauge symmetry via the D- and F-flatness preserving Higgs mechanism in which 
Higgs fields are the massless open string states from a specific $N=2$ subsector. 
Moreover,  Models \ref{twoOplane1} and \ref{twoOplane01} are 
T-dual to each other and can be obtained
from Model I-Z-2 in \cite{Cvetic:2004ui} via generic Type II T-duality transformation.  
Thus, the $SU(3)_C \times SU(2)_L$ gauge coupling unification in Model I-Z-2 
is shifted to $SU(3)_C \times U(1)_Y$ gauge coupling unification in Models \ref{twoOplane1} and \ref{twoOplane01},
and vice versa. Also,
Model \ref{threeOplane1} with $b$- and $c$-stack of D6-branes swapped 
leads to Model \ref{threeOplane01} with  $SU(4)$ and $SU(2)_L$ gauge couplings being closer to unification
at the string scale. 

Furthermore, we obtained the models with one $USp(4)$ group, and 
the new confine particle spectrum in Model \ref{oneOplane02}
shown in Table \ref{Composite Particles oneOplane02}. The composite particle spectrum of Model \ref{twoOplane01}, 
which is formed due to  two confined $USp(4)$ groups in hidden sector, is given 
in Table \ref{Composite Particles twoOplane01}, where supersymmetry breaking  
via a ``race-track'' scenario is still possible.

Last but not least, we found interesting models with wrapping number larger than or equal to $5$ in supervised scanning methods, which was not found before. For these models, especially Model \ref{5wrap2} and its T-duality model, the gauge couplings are in a much more refined form because of the complicity of the intersections due to a large wrapping number. An approximate gauge coupling 
unification is achieved. For the models with larger wrapping numbers, a better gauge unified model was not found yet for large wrapping number up to $10$. However, due to the limited scanning so far, it is worthwhile to search for the Pati-Salam models 
with broader scanning and large wrapping numbers. And in the machine learning and model building part,  we observed visually that the MSSM-like models tend to gather in islands, which indicates more chances to find more MSSM-like models  in their nearby scanning region.

\section*{Acknowledgments} 

This research was supported  by the Projects 11847612 and 11875062 supported by the 
National Natural Science Foundation of China, the Key Research Program of Frontier Science, CAS, 
the National Thousand Young Talents Program of China, the China Postdoctoral Science Foundation Grant 2018M631436, 
and the LMU Munich's Institutional Strategy LMUexcellent within the framework of the German Excellence Initiative. 
We would like to thank   Jie Ren and Zheng Sun for discussing and participating for the early part of the work,
thank Xiaoyong Chu, Andreas Deser, Jiahua Tian, and Yinan Wang for useful discussions. 
RS would also like to thank Babak Haghighat for his support, and acknowledges Ludwig Maximilian University of Munich, Max Planck Institute for Physics, the Abdus Salam International Centre for Theoretical Physics~(ICTP) for their hospitalities where part of this work was carried out.

\begin{appendix}
\begin{center}
\Large{\bf Appendix: Supersymmetric Pati-Salam Models}
\end{center}

In this Appendix,  we tabulate $12$ representative models obtained from our broader scanning method.  
In the first column for each table,
we denote the $U(4)$, $U(2)_L$, and $U(2)_R$ stacks of D6-branes as  $a$, $b$, and $c$ stacks,
respectively. We also employ 1, 2, 3, and 4 stacks to represent the filler branes
respectively along $\Omega R$, $\Omega R\omega$, $\Omega R\theta\omega$, and 
$\Omega R\theta$ orientifold planes, which result in the $USp(N)$ gauge symmetries.  
In the second column, $N$ is the number of D6-branes in each stack. Moreover, 
we present the wrapping numbers of the various D6-branes in the third column
and specify the third set of wrapping numbers for the tilted two-torus. 

In the remaining right columns, we give the intersection numbers between 
various stacks, where $b'$ and $c'$ denote the $\Omega R$
images of $b$ and $c$, respectively. 
In addition, we present the relation among the  moduli parameters imposed 
by the four-dimensional $N=1$ supersymmetry conditions, 
and the one-loop $\beta$ functions ($\beta^g_i$) for the hidden sector gauge
symmetries. In particular, we also give the MSSM gauge couplings in the caption of
each model, and thus it is easier to check the gauge coupling unification.
 However,  we do not require  at least two confining hidden gauge sectors
which are needed to realize the moduli stabilization and supersymmetry breaking via gaugino condensation.

\begin{table}[ht]
	\footnotesize
	\renewcommand{\arraystretch}{1.0}
	\caption{D6-brane configurations and intersection numbers in Model \ref{oneOplane1},
and its MSSM gauge coupling relation is
 $g^2_a=\frac{1}{2}\, g^2_b=\frac{65}{44} \,(\frac{5}{3}\,g^2_Y)= \frac{16 \sqrt{6}}{15} \, \pi \,e^{\phi^4}$.}
		\label{oneOplane1}
	\begin{center}
		\begin{tabular}{|c||c|c||c|c|c|c|c|c|c|}
			\hline	 \rm{Model} \ref{oneOplane1} &\multicolumn{9}{c|}{$U(4)\times U(2)_L\times U(2)_R\times USp(4)$}\\
			\hline \hline \rm{stack} & $N$ & $(n^1,l^1)\times (n^2,l^2)\times
			(n^3,l^3)$ & $n_{\Ysymm}$& $n_{\Yasymm}$ & $b$ & $b'$ & $c$ & $c'$& 1   \\
			\hline
			$a$&  8& $(1,-2)\times (0,1)\times (1,-1)$ & 1 & -1  & 3 & 0 & -3 & 0 & -2  \\
			$b$&  4& $(0,1)\times (-3,-1)\times (1,1)$ & 2 & -2  & - & - & 0 & -4     & 1     \\
			$c$&  4& $(2,-1)\times (-1,1)\times (-1,-1)$ & -2 & -6  & - & - & - & - & -1   \\
			\hline
			1&   4& $(1,0)\times (1,0)\times (2,0)$& \multicolumn{7}{c|}{$X_A = 3 X_B =\frac{1}{2}X_C =\frac{3}{8}X_D$}\\
			&      &             & \multicolumn{7}{c|}{$\beta^g_1=0$;\quad $\chi_1=\frac{1}{4},\chi_2=\frac{3}{2},\chi_3=4$}\\
			\hline
		\end{tabular}
	\end{center}
\end{table}

\begin{table}[ht]
	\footnotesize
	\renewcommand{\arraystretch}{1.0}
	\caption{ D6-brane configurations and intersection numbers in Model \ref{oneOplane01},
and its MSSM gauge coupling relation is
$g^2_a=\frac{1}{2}\, g^2_b=\frac{65}{44} \,(\frac{5}{3}\,g^2_Y)= \frac{16 \sqrt{6}}{15} \, \pi \,e^{\phi^4}$.}
	\label{oneOplane01}
	\begin{center}
		\begin{tabular}{|c||c|c||c|c|c|c|c|c|c|}
				\hline	 \rm{Model} \ref{oneOplane01}&\multicolumn{9}{|c|}{$U(4)\times U(2)_L\times U(2)_R\times USp(4)$}\\
			\hline \hline \rm{stack} & $N$ & $(n^1,l^1)\times (n^2,l^2)\times
			(n^3,l^3)$ & $n_{\Ysymm}$& $n_{\Yasymm}$ & $b$ & $b'$ & $c$ & $c'$& 2   \\
			\hline
			$a$&  8& $(-1,2)\times (-1,0)\times (1,1)$ & 1 & -1  & 0 & 3 & -3 & 0 & -2  \\
			$b$&  4& $(0,1)\times (-1,-3)\times (1,1)$ & -2 & 2  & - & - & 4 & 0     & -1     \\
			$c$&  4& $(2,-1)\times (-1,-1)\times (-1,1)$ & -2 & -6  & - & - & - & - & -1   \\
			\hline
			2&   4& $(1,0)\times (0,-1)\times (0,2)$& \multicolumn{7}{c|}{$X_A = \frac{1}{3} X_B =\frac{1}{8}X_C =\frac{1}{6}X_D$}\\
			&      &             & \multicolumn{7}{c|}{$\beta^g_2=0$;\quad $\chi_1=\frac{1}{4},\chi_2=\frac{2}{3},\chi_3=1$}\\
			\hline
		\end{tabular}
	\end{center}
\end{table}

\begin{table}[ht]
	\footnotesize
	\renewcommand{\arraystretch}{1.0}
	\caption{ D6-brane configurations and intersection numbers in Model \ref{oneOplane2},
and its MSSM gauge coupling relation is
$g^2_a=2\, g^2_b=\frac{10}{13} \,(\frac{5}{3}\,g^2_Y)= \frac{16 \sqrt{6}}{15} \, \pi \,e^{\phi^4}$.}
		\label{oneOplane2}
	\begin{center}
		\begin{tabular}{|c||c|c||c|c|c|c|c|c|c|}
			\hline	 \rm{Model} \ref{oneOplane2}&
			\multicolumn{9}{|c|}{$U(4)\times U(2)_L\times U(2)_R\times USp(4)$}\\
			\hline \hline \rm{stack} & $N$ & $(n^1,l^1)\times (n^2,l^2)\times
			(n^3,l^3)$ & $n_{\Ysymm}$& $n_{\Yasymm}$ & $b$ & $b'$ & $c$ & $c'$& 2   \\
			\hline
			$a$&  8& $(1,-1)\times (0,-1)\times (-1,1)$ & 0 & 0  & 3 & 0 & -3 & 0 & 0  \\
			$b$&  4& $(0,-1)\times (3,2)\times (1,1)$ & 1 & -1  & - & - & 0 & 8     & -3     \\
			$c$&  4& $(4,-1)\times (-1,0)\times (-1,-1)$ & -3 & 3  & - & - & - & - & -1   \\
			\hline
			2&   4& $(1,0)\times (0,-1)\times (0,2)$& \multicolumn{7}{c|}{$X_A = \frac{3}{2}X_B =X_C =6X_D$}\\
			&      &                                                & \multicolumn{7}{c|}{$\beta^g_2=-2$;\quad $\chi_1=2,\chi_2=3,\chi_3=1$}\\
			\hline
		\end{tabular}
	\end{center}
\end{table}

\begin{table}[ht]
	\footnotesize
	\renewcommand{\arraystretch}{1.0}
	\caption{D6-brane configurations and intersection numbers in Model \ref{oneOplane02},
and its MSSM gauge coupling relation is
$g^2_a=\frac{2}{3}\, g^2_b=\frac{10}{7} \,(\frac{5}{3}\,g^2_Y)= \frac{16 \sqrt{6}}{15} \, \pi \,e^{\phi^4}$.}
	\label{oneOplane02}
	\begin{center}
		\begin{tabular}{|c||c|c||c|c|c|c|c|c|c|}
				\hline	 \rm{Model} \ref{oneOplane02}&
			\multicolumn{9}{|c|}{$U(4)\times U(2)_L\times U(2)_R\times USp(4)$}\\
			\hline \hline \rm{stack} & $N$ & $(n^1,l^1)\times (n^2,l^2)\times
			(n^3,l^3)$ & $n_{\Ysymm}$& $n_{\Yasymm}$ & $b$ & $b'$ & $c$ & $c'$& 2   \\
			\hline
			$a$&  8& $(-1,0)\times (1,-1)\times (-1,-1)$ & 0 & 0  & 3 & 0 & 0 & -3 & 0  \\
			$b$&  4& $(0,1)\times (1,-4)\times (1,-1)$ & 3 & -3  & - & - & -8 & 0    & 1     \\
			$c$&  4& $(2,-3)\times (1,0)\times (1,1)$ & 1 & -1  & - & - & - & - & -3   \\
			\hline
			2&   4& $(1,0)\times (0,-1)\times (0,2)$& \multicolumn{7}{c|}{$X_A = \frac{1}{4}X_B =\frac{1}{6}X_C =\frac{1}{6}X_D$}\\
			&      &                                                & \multicolumn{7}{c|}{$\beta^g_2=-2$;\quad $\chi_1=\frac{1}{3}, \chi_2=\frac{1}{2}, \chi_3=1$}\\
			\hline
		\end{tabular}
	\end{center}
\end{table}

\begin{table}[ht]
	\footnotesize
	\renewcommand{\arraystretch}{1.0}
	\caption{D6-brane configurations and intersection numbers in Model \ref{twoOplane1},
and its MSSM gauge coupling relation is
$g^2_a=\frac{5}{3}\, g^2_b=\frac{5}{3}\,g^2_Y= \frac{4 \sqrt{6}}{3}\, \pi \,e^{\phi^4}$.}
	\label{twoOplane1}
	\begin{center}
		\begin{tabular}{|c||c|c||c|c|c|c|c|c|c|c|}
				\hline	 \rm{Model} \ref{twoOplane1}&
			\multicolumn{10}{|c|}{$U(4)\times U(2)_L\times U(2)_R\times USp(4)^2$}\\
			\hline \hline \rm{stack} & $N$ & $(n^1,l^1)\times (n^2,l^2)\times
			(n^3,l^3)$ & $n_{\Ysymm}$& $n_{\Yasymm}$ & $b$ & $b'$ & $c$ & $c'$& 1 & 2  \\
			\hline
			$a$&  8& $(0,-1)\times (1,1)\times (1,1)$ & 0 & 0  & 3 & 0 & -3 & 0 & 1  & -1  \\
			$b$&  4& $(1,1)\times (2,-1)\times (1,-1)$ & -2 & -6  & - & - & 0 & 4 & -1 &2   \\
			$c$&  4& $(-3,1)\times (0,-1)\times (1,-1)$ & -2 & 2  & - & - & - & - & -1  & 0  \\
			\hline
			1&   4& $(1,0)\times (1,0)\times (2,0)$& \multicolumn{8}{c|}{$X_A = X_B =3 X_C =3X_D$}\\
			2&   4& $ (1,0)\times(0,-1)\times (0,2)$ & \multicolumn{8}{c|}{$\beta^g_1=-2, \beta^g_2=-2$;\quad $\chi_1=3, \chi_2=1, \chi_3=2$}\\
			\hline
		\end{tabular}
	\end{center}
\end{table}

\begin{table}[ht]
	\footnotesize
	\renewcommand{\arraystretch}{1.0}
	\caption{D6-brane configurations and intersection numbers in Model \ref{twoOplane01},
and its MSSM gauge coupling relation is
$g^2_a=\frac{5}{3}\, g^2_b=\frac{5}{3}\,g^2_Y= \frac{4 \sqrt{6}}{3}\, \pi \,e^{\phi^4}$.}
	\label{twoOplane01}
	\begin{center}
		\begin{tabular}{|c||c|c||c|c|c|c|c|c|c|c|}
				\hline	 \rm{Model} \ref{twoOplane01}&
			\multicolumn{10}{|c|}{$U(4)\times U(2)_L\times U(2)_R\times USp(4)^2$}\\
			\hline \hline \rm{stack} & $N$ & $(n^1,l^1)\times (n^2,l^2)\times
			(n^3,l^3)$ & $n_{\Ysymm}$& $n_{\Yasymm}$ & $b$ & $b'$ & $c$ & $c'$& 3 & 4  \\
			\hline
			$a$&  8& $(-1,0)\times (1,-1)\times (-1,-1)$ & 0 & 0  & 3 & 0 & 0 & -3 & -1  & 1  \\
			$b$&  4& $(-1,1)\times (1,2)\times (-1,1)$ & -2 & -6  & - & - & 4 & 0 & 2 &-1   \\
			$c$&  4& $(1,-3)\times (-1,0)\times (-1,-1)$ & 2 & -2  & - & - & - & - & 0  & 1  \\
			\hline
			3&   4& $(0,-1)\times (1,0)\times (0,2)$& \multicolumn{8}{c|}{$X_A = X_B =\frac{1}{3} X_C =\frac{1}{3}X_D$}\\
			4&   4& $ (0,-1)\times(0,1)\times (2,0)$ & \multicolumn{8}{c|}{$\beta^g_3=-2, \beta^g_4=-2$;\quad $\chi_1=\frac{1}{3}, \chi_2=1, \chi_3=2$}\\
			\hline
		\end{tabular}
	\end{center}
\end{table}

\begin{table}[ht]
	\footnotesize
	\renewcommand{\arraystretch}{1.0}
	\caption{D6-brane configurations and intersection numbers in Model \ref{threeOplane1},
and its MSSM gauge coupling relation is
$g^2_a=\frac{2}{3}\, g^2_b=\frac{80}{77} \,(\frac{5}{3}\,g^2_Y)= \frac{16 \sqrt{6}}{15} \, \pi \,e^{\phi^4}$.}
	\label{threeOplane1}
	\begin{center}
		\begin{tabular}{|c||c|c||c|c|c|c|c|c|c|c|c|}
				\hline	 \rm{Model} \ref{threeOplane1}&
			\multicolumn{11}{|c|}{$U(4)\times U(2)_L\times U(2)_R\times USp(2)\times USp(4)\times USp(4)$}\\
			\hline \hline \rm{stack} & $N$ & $(n^1,l^1)\times (n^2,l^2)\times
			(n^3,l^3)$ & $n_{\Ysymm}$& $n_{\Yasymm}$ & $b$ & $b'$ & $c$ & $c'$& 1 & 2 & 4  \\
			\hline
			$a$&  8& $(-1,1)\times (-1,0)\times (1,1)$ & 0 & 0  & 0 & 3 & -3 & 0 & 0 & -1 & 1  \\
			$b$&  4& $(1,4)\times (0,-1)\times (1,1)$ & -3 & 3  & - & - & 2 & 0 & 4 & 0 &0   \\
			$c$&  4& $(-1,-2)\times (-1,-1)\times (1,-1)$ & 0 & 8  & - & - & - & - & 2 & 2 & -1  \\
			\hline
			1&   2& $(1,0)\times (1,0)\times (2,0)$& \multicolumn{9}{c|}{$X_A =\frac{1}{6} X_B =\frac{1}{4}X_C =\frac{1}{6}X_D$}\\
			2&   4& $(1,0)\times (0,-1)\times (0,2)$ & \multicolumn{9}{c|}{$\beta^g_1=0, \beta^g_2=-2, \beta^g_4=-3$}\\
			4&  4 & $(0,-1)\times (0,1)\times (2,0)$ & \multicolumn{9}{c|}{$\chi_1=\frac{1}{2},\chi_2=\frac{1}{3},\chi_3=1$}\\
			\hline
		\end{tabular}
	\end{center}
\end{table}

\begin{table}[ht]
	\footnotesize
	\renewcommand{\arraystretch}{1.0}
	\caption{D6-brane configurations and intersection numbers in Model \ref{threeOplane01},
and its MSSM gauge coupling relation is
$g^2_a=\frac{16}{15}\, g^2_b=\frac{10}{13} \,(\frac{5}{3}\,g^2_Y)= \frac{16 \sqrt{6}}{15} \, \pi \,e^{\phi^4}$.}
	\label{threeOplane01}
	\begin{center}
		\begin{tabular}{|c||c|c||c|c|c|c|c|c|c|c|c|}
			\hline	 \rm{Model} \ref{threeOplane01}&
			\multicolumn{11}{|c|}{$U(4)\times U(2)_L\times U(2)_R\times USp(2)\times USp(4)\times USp(4)$}\\
			\hline \hline \rm{stack} & $N$ & $(n^1,l^1)\times (n^2,l^2)\times
			(n^3,l^3)$ & $n_{\Ysymm}$& $n_{\Yasymm}$ & $b$ & $b'$ & $c$ & $c'$& 1 & 3 & 4  \\
			\hline
			$a$&  8& $(1,-1)\times (0,1)\times (1,-1)$ & 0 & 0  & 0 & 3 & -3 & 0 & -1 & 1 & 0  \\
			$b$&  4& $(-2,1)\times (1,-1)\times (1,-1)$ & 0 & 8  & - & - & -2 & 0 & -1 & 2 &2   \\
			$c$&  4& $(-4,1)\times (1,0)\times (-1,-1)$ & -3 & 3  & - & - & - & - & 0 & 0 & 4  \\
			\hline
			1&   4& $(1,0)\times (1,0)\times (2,0)$& \multicolumn{9}{c|}{$X_A =\frac{3}{2} X_B = X_C =6 X_D$}\\
			3&   4& $(0,-1)\times (1,0)\times (0,2)$ & \multicolumn{9}{c|}{$\beta^g_1=-3, \beta^g_3=-2, \beta^g_4=0$}\\
			4&  2 & $(0,-1)\times (0,1)\times (2,0)$ & \multicolumn{9}{c|}{$\chi_1= 2,\chi_2= 3, \chi_3=1$}\\
			\hline
		\end{tabular}
	\end{center}
\end{table}

\begin{table}[ht]
	\footnotesize
	\renewcommand{\arraystretch}{1.0}
	\caption{D6-brane configurations and intersection numbers in Model \ref{fourOplane1},
and its MSSM gauge coupling relation is
$g^2_a= g^2_b=\frac{10}{13} \,(\frac{5}{3}\,g^2_Y)= \frac{16\times 2^{1/4}}{3\sqrt{3}} \, \pi \,e^{\phi^4}$.}
	\label{fourOplane1}
	\begin{center}
		\begin{tabular}{|c||c|c||c|c|c|c|c|c|c|c|c|c|}
			\hline	 \rm{Model} \ref{fourOplane1}&
			\multicolumn{12}{|c|}{$U(4)\times U(2)_L\times U(2)_R\times USp(2)\times USp(4)\times USp(2)\times USp(2)$}\\
			\hline \hline \rm{stack} & $N$ & $(n^1,l^1)\times (n^2,l^2)\times
			(n^3,l^3)$ & $n_{\Ysymm}$& $n_{\Yasymm}$ & $b$ & $b'$ & $c$ & $c'$& 1 & 2 & 3 & 4 \\
			\hline
			$a$&  8& $(1,0)\times (1,-1)\times (1,1)$ & 0 & 0  & 3 & 0 & 0 & -3 & 0 & 0 & -1 & 1  \\
			$b$&  4& $(1,3)\times (-1,2)\times (-1,1)$ & -6 & -18  & - & - & 9 & 0 & -6 & 3 &-2 & -1   \\
			$c$&  4& $(-2,3)\times (0,1)\times (1,1)$ & 5 & -5  & - & - & - & - & -3 & 0 & -2 & 0  \\
			\hline
			1&   2& $(1,0)\times (1,0)\times (2,0)$& \multicolumn{10}{c|}{$X_A = 2X_B =\frac{2}{3}X_C =\frac{2}{3}X_D$}\\
			2&   4& $(1,0)\times (0,-1)\times (0,2)$ & \multicolumn{10}{c|}{$\beta^g_1=3, \beta^g_2=-3, \beta^g_3=0, \beta^g_4=-3$}\\
			3&  2 & $(0,-1)\times (1,0)\times (0,2)$ & \multicolumn{10}{c|}{$\chi_1=\frac{\sqrt{2}}{{3}},\chi_2=\sqrt{2}, \chi_3=2\sqrt{2}$}\\
			4&  2 & $(0,-1)\times (0,1)\times (2,0)$ & \multicolumn{10}{c|}{}\\
			\hline
		\end{tabular}
	\end{center}
\end{table}

\begin{table}[ht]
	\footnotesize
	\renewcommand{\arraystretch}{1.0}
	\caption{D6-brane configurations and intersection numbers in Model \ref{fourOplaneT},
and its MSSM gauge coupling relation is
$g^2_a= g^2_b=\frac{5}{3}\,g^2_Y= \frac{{4}\sqrt{6}}{3} \, \pi \,e^{\phi^4}$.}
		\label{fourOplaneT}
	\begin{center}
		\begin{tabular}{|c||c|c||c|c|c|c|c|c|c|c|c|c|}
				\hline	 \rm{Model} \ref{fourOplaneT}&
			\multicolumn{12}{|c|}{$U(4)\times U(2)_L\times U(2)_R\times USp(2)^2$}\\
			\hline \hline \rm{stack} & $N$ & $(n^1,l^1)\times (n^2,l^2)\times
			(n^3,l^3)$ & $n_{\Ysymm}$& $n_{\Yasymm}$ & $b$ & $b'$ & $c$ & $c'$& 1 & 2 & 3 & 4 \\
			\hline
			$a$&  8& $(1,0)\times (1,-1)\times (1,1)$ & 0 & 0  & 0 & 3 & 0 & -3 & 0 & 0 & -1 & 1  \\
			$b$&  4& $(-1,-3)\times (0,-1)\times (-1,-1)$ & -2 & 2  & - & - & 0 & 0 & 3 & 0 &-1 & 0   \\
			$c$&  4& $(1,-3)\times (-1,0)\times (-1,-1)$ & 2 & -2  & - & - & - & - & 0 & -3 & 0 & 1  \\
			\hline
			1&   2& $(1,0)\times (1,0)\times (2,0)$& \multicolumn{10}{c|}{$3X_A = 3X_B =X_C =X_D$}\\
			2&   2& $(1,0)\times (0,-1)\times (0,2)$ & \multicolumn{10}{c|}{$\beta^g_1=-3, \beta^g_2=-3, \beta^g_3=-3, \beta^g_4=-3$}\\
			3&  2 & $(0,-1)\times (1,0)\times (0,2)$ & \multicolumn{10}{c|}{$\chi_1=\frac{1}{{3}},\chi_2=1, \chi_3=2$}\\
			4&  2 & $(0,-1)\times (0,1)\times (2,0)$ & \multicolumn{10}{c|}{}\\
			\hline
		\end{tabular}
	\end{center}
\end{table}

\begin{table}[ht]
	\footnotesize
	\renewcommand{\arraystretch}{1.0}
	\caption{D6-brane configurations and intersection numbers in Model \ref{5wrap0}, and its MSSM gauge coupling relation is $g^2_a=\frac{5}{6}\, g^2_b=\frac{11}{8} \,(\frac{5}{3}\,g^2_Y)= \frac{8\times 2^{1/4}\times 5^{3/4}}{7\sqrt{3}} \, \pi \,e^{\phi^4}$.}
	\label{5wrap0}
	\begin{center}
		\begin{tabular}{|c||c|c||c|c|c|c|c|c|c|}
			\hline	 \rm{Model} \ref{5wrap0}&
			\multicolumn{9}{c|}{$U(4)\times U(2)_L\times U(2)_R\times USp(2)$}\\
			\hline \hline \rm{stack} & $N$ & $(n^1,l^1)\times (n^2,l^2)\times
			(n^3,l^3)$ & $n_{\Ysymm}$& $n_{\Yasymm}$ & $b$ & $b'$ & $c$ & $c'$ & 3 \\
			\hline
			$a$&  8& $(0,1)\times (1,1)\times (-1,-1)$ & 0 & 0  & 0 & 3 & 0 & -3 & 0  \\
			$b$&  4& $(1,0)\times (5,-2)\times (1,1)$ & -3 & 3  & - & - & 0 & 8  & -2   \\
			$c$&  4& $(-1,-1)\times (1,2)\times (1,1)$ & -2 & -6  & - & - & - & -  & -2  \\
			\hline
			3&   2& $(0,-1)\times (1,0)\times (0,2)$& \multicolumn{7}{c|}{$X_A =  X_B =\frac{12}{5}X_C = 6 X_D$}\\
			&      &             & \multicolumn{7}{c|}{$\beta^g_3=-2$;\quad $\chi_1=\frac{6\sqrt{10}}{5} ,\chi_2=\frac{\sqrt{10}}{2},\chi_3=\frac{2\sqrt{10}}{5}$}\\
			\hline
		\end{tabular}
	\end{center}
\end{table}

\begin{table}[ht]
	\footnotesize
	\renewcommand{\arraystretch}{1.0}
	\caption{D6-brane configurations and intersection numbers in Model \ref{5wrap1},
and its MSSM gauge coupling relation is
$g^2_a=\frac{5}{6}\, g^2_b=\frac{35}{32} \,(\frac{5}{3}\,g^2_Y)= \frac{8\times 5^{3/4}\sqrt{7}}{27} \, \pi \,e^{\phi^4}$.}
	\label{5wrap1}
	\begin{center}
		\begin{tabular}{|c||c|c||c|c|c|c|c|c|c|c|}
			\hline	 \rm{Model} \ref{5wrap1}&
			\multicolumn{10}{|c|}{$U(4)\times U(2)_L\times U(2)_R\times USp(2)^2$}\\
			\hline \hline \rm{stack} & $N$ & $(n^1,l^1)\times (n^2,l^2)\times
			(n^3,l^3)$ & $n_{\Ysymm}$& $n_{\Yasymm}$ & $b$ & $b'$ & $c$ & $c'$& 2 & 3  \\
			\hline
			$a$&  8& $(1,-1)\times (1,1)\times (1,-1)$ & 0 & -4  & 3 & 0 & -3 & 0 & -1 & 1  \\
			$b$&  4& $(-2,5)\times (-1,0)\times (1,1)$ & 3 & -3  & - & - & 0 & -1 & -5 & 0   \\
			$c$&  4& $(-1,-2)\times (0,-1)\times (-1,-1)$ & -1 & 1  & - & - & - & - & 0 & -1  \\
			\hline
			2&   2& $(1,0)\times (0,-1)\times (0,2)$& \multicolumn{8}{c|}{$X_A = \frac{7}{2}X_B =\frac{1}{2}X_C =\frac{7}{5}X_D$}\\
			3&   2& $(0,-1)\times (1,0)\times (0,2)$ & \multicolumn{8}{c|}{$\beta^g_2=1, \beta^g_3=-3$}\\
			&    &                                                  & \multicolumn{8}{c|}{$\chi_1=\frac{\sqrt{5}}{5},\chi_2=\frac{7\sqrt{5}}{5},\chi_3=\sqrt{5}$}\\
			\hline
		\end{tabular}
	\end{center}
\end{table}

\begin{table}[ht]
	\footnotesize
	\renewcommand{\arraystretch}{1.0}
	\caption{D6-brane configurations and intersection numbers in Model \ref{5wrap2},
and its MSSM gauge coupling relation is
$g^2_a=\frac{71}{63} g^2_b=\frac{50}{47} \,(\frac{5}{3}\,g^2_Y)= \frac{16\times 2^{1/4}\times 5^{3/4}}{21}  \pi e^{\phi^4}$.}
		\label{5wrap2}
	\begin{center}
		\begin{tabular}{|c||c|c||c|c|c|c|c|c|c|c|c|}
			\hline	 \rm{Model} \ref{5wrap2}&
			\multicolumn{11}{|c|}{$U(4)\times U(2)_L\times U(2)_R\times USp(2)^3$}\\
			\hline \hline \rm{stack} & $N$ & $(n^1,l^1)\times (n^2,l^2)\times
			(n^3,l^3)$ & $n_{\Ysymm}$& $n_{\Yasymm}$ & $b$ & $b'$ & $c$ & $c'$& 2 & 3 & 4  \\
			\hline
			$a$&  8& $(-1,1)\times (-1,0)\times (1,1)$ & 0 & 0  & 0 & 3 & -3 & 0 & -1 & 0 & 1  \\
			$b$&  4& $(2,-1)\times (1,1)\times (1,1)$ & 0 & 8  & - & - & -1 & 0 & -1 & 2 &2   \\
			$c$&  4& $(5,-2)\times (0,1)\times (1,-1)$ & -3 & 3  & - & - & - & - & 0 & 5 & 0  \\
			\hline
			2&   2& $(1,0)\times (0,-1)\times (0,2)$& \multicolumn{9}{c|}{$X_A =\frac{5}{9} X_B =\frac{5}{2}X_C =\frac{5}{9}X_D$}\\
			3&   2& $(0,-1)\times (1,0)\times (0,2)$ & \multicolumn{9}{c|}{$\beta^g_2=-3, \beta^g_3=1, \beta^g_4=-2$}\\
			4&  2 & $(0,-1)\times (0,1)\times (2,0)$ & \multicolumn{9}{c|}{$\chi_1=\frac{\sqrt{10}}{2},\chi_2=\frac{\sqrt{10}}{9},\chi_3=\sqrt{10}$}\\
			\hline
		\end{tabular}
	\end{center}
\end{table}

\begin{table}[ht]
	\footnotesize
	\renewcommand{\arraystretch}{1.0}
	\caption{D6-brane configurations and intersection numbers in Model \ref{5wrap01}, and its MSSM gauge coupling relation is $g^2_a=\frac{11}{6}\, g^2_b=\frac{25}{28} \,(\frac{5}{3}\,g^2_Y)= \frac{8\times 2^{1/4}\times 5^{3/4}}{7\sqrt{3}} \, \pi \,e^{\phi^4}$. }
	\label{5wrap01}
	\begin{center}
		\begin{tabular}{|c||c|c||c|c|c|c|c|c|c|}
			\hline	 \rm{Model} \ref{5wrap01}&
			\multicolumn{9}{c|}{$U(4)\times U(2)_L\times U(2)_R\times USp(2)$}\\
			\hline \hline \rm{stack} & $N$ & $(n^1,l^1)\times (n^2,l^2)\times
			(n^3,l^3)$ & $n_{\Ysymm}$& $n_{\Yasymm}$ & $b$ & $b'$ & $c$ & $c'$ & 3 \\
			\hline
			$a$&  8& $(0,-1)\times (1,-1)\times (-1,1)$ & 0 & 0  & 0 & 3 & 0 & -3 & 0  \\
			$b$&  4& $(-1,1)\times (1,-2)\times (1,-1)$ & 2 & 6  & - & - & 0 & -8  & 2  \\
			$c$&  4& $(1,0)\times (5,2)\times (1,-1)$ & 3 & -3  & - & - & - & -  & 2   \\
			\hline
			3&   2& $(0,-1)\times (1,0)\times (0,2)$& \multicolumn{7}{c|}{$X_A =  X_B =\frac{12}{5}X_C = 6 X_D$}\\
			&      &             & \multicolumn{7}{c|}{$\beta^g_3=-2$;\quad $\chi_1=\frac{6\sqrt{10}}{5} ,\chi_2=\frac{\sqrt{10}}{2},\chi_3=\frac{2\sqrt{10}}{5}$}\\
			\hline
		\end{tabular}
	\end{center}
\end{table}

\begin{table}[ht]
	\footnotesize
	\renewcommand{\arraystretch}{1.0}
	\caption{D6-brane configurations and intersection numbers in Model \ref{5wrap11},
		and its MSSM gauge coupling relation is $g^2_a=\frac{7}{6}\, g^2_b=\frac{25}{28} \,(\frac{5}{3}\,g^2_Y)= \frac{8\times 5^{3/4}\sqrt{7}}{27} \, \pi \,e^{\phi^4}$. }
	\label{5wrap11}
	\begin{center}
		\begin{tabular}{|c||c|c||c|c|c|c|c|c|c|c|}
			\hline	 \rm{Model} \ref{5wrap11}&
			\multicolumn{10}{|c|}{$U(4)\times U(2)_L\times U(2)_R\times USp(2)^2$}\\
			\hline \hline \rm{stack} & $N$ & $(n^1,l^1)\times (n^2,l^2)\times
			(n^3,l^3)$ & $n_{\Ysymm}$& $n_{\Yasymm}$ & $b$ & $b'$ & $c$ & $c'$& 2 & 3  \\
			\hline
			$a$&  8& $(1,1)\times (1,-1)\times (1,1)$ & 0 & 4  & 3 & 0 & -3 & 0 & 1 & -1  \\
			$b$&  4& $(-1,2)\times (0,1)\times (-1,1)$ & 1 & -1  & - & - & 0 & 1 & 0 & 1  \\
			$c$&  4& $(-2,-5)\times (-1,0)\times (1,-1)$ & -3 & 3  & - & - & - & - & 5 & 0   \\
			\hline
			2&   2& $(1,0)\times (0,-1)\times (0,2)$& \multicolumn{8}{c|}{$X_A = \frac{7}{2}X_B =\frac{1}{2}X_C =\frac{7}{5}X_D$}\\
			3&   2& $(0,-1)\times (1,0)\times (0,2)$ & \multicolumn{8}{c|}{$\beta^g_2=1, \beta^g_3=-3$}\\
			&    &                                                  & \multicolumn{8}{c|}{$\chi_1=\frac{\sqrt{5}}{5},\chi_2=\frac{7\sqrt{5}}{5},\chi_3=\sqrt{5}$}\\
			\hline
		\end{tabular}
	\end{center}
\end{table}

\begin{table}[ht]
	\footnotesize
	\renewcommand{\arraystretch}{1.0}
	\caption{D6-brane configurations and intersection numbers in Model \ref{5wrap21},
		and its MSSM gauge coupling relation is $g^2_a=\frac{10}{9} g^2_b=\frac{355}{331} \,(\frac{5}{3}\,g^2_Y)= \frac{16\times 2^{1/4}\times 5^{3/4}}{21}  \pi e^{\phi^4}$.}
	\label{5wrap21}
	\begin{center}
		\begin{tabular}{|c||c|c||c|c|c|c|c|c|c|c|c|}
			\hline	 \rm{Model} \ref{5wrap21}&
			\multicolumn{11}{|c|}{$U(4)\times U(2)_L\times U(2)_R\times USp(2)^3$}\\
			\hline \hline \rm{stack} & $N$ & $(n^1,l^1)\times (n^2,l^2)\times
			(n^3,l^3)$ & $n_{\Ysymm}$& $n_{\Yasymm}$ & $b$ & $b'$ & $c$ & $c'$& 2 & 3 & 4  \\
			\hline
			$a$&  8& $(-1,-1)\times (-1,0)\times (1,-1)$ & 0 & 0  & 0 & 3 & -3 & 0 & 1 & 0 & -1  \\
			$b$&  4& $(5,2)\times (0,-1)\times (1,1)$ & 3 & -3  & - & - & -1 & 0 & 0 & -5 & 0  \\
			$c$&  4& $(2,1)\times (1,-1)\times (1,-1)$ & 0 & -8  & - & - & - & - & 1 & -2 &-2   \\
			\hline
			2&   2& $(1,0)\times (0,-1)\times (0,2)$& \multicolumn{9}{c|}{$X_A =\frac{5}{9} X_B =\frac{5}{2}X_C =\frac{5}{9}X_D$}\\
			3&   2& $(0,-1)\times (1,0)\times (0,2)$ & \multicolumn{9}{c|}{$\beta^g_2=-3, \beta^g_3=1, \beta^g_4=-2$}\\
			4&  2 & $(0,-1)\times (0,1)\times (2,0)$ & \multicolumn{9}{c|}{$\chi_1=\frac{\sqrt{10}}{2},\chi_2=\frac{\sqrt{10}}{9},\chi_3=\sqrt{10}$}\\
			\hline
		\end{tabular}
	\end{center}
\end{table}


\begin{table}[ht]
	\footnotesize
	\renewcommand{\arraystretch}{1.0}
	\caption{D6-brane configurations and intersection numbers of Model \ref{model5}, and its MSSM gauge coupling relation is $g^2_a=\frac{7219}{4179}g^2_b=\frac{275}{1901}(\frac{5}{3}g^2_Y)=\frac{32\times 3^{1/4}85^{3/4}}{1393}\pi e^{\phi^4}$.} \label{model5}
	\begin{center}
		\begin{tabular}{|c||c|c||c|c|c|c|c|c|c|c|c|c|}
			\hline
			\rm{Model} \ref{model5} & \multicolumn{12}{c|}{$U(4)\times U(2)_L\times U(2)_R\times USp(12)\times USp(2)$}\\
			\hline \hline \rm{stack} & $N$ & $(n^1,l^1)\times (n^2,l^2)\times
			(n^3,l^3)$ & $n_{\Ysymm}$& $n_{\Yasymm}$ & $b$ & $b'$ & $c$ & $c'$& 1 & 2 & 3 & 4 \\
			\hline
			$a$&  8& $(-1,1)\times (-6,1)\times (-1,1)$ & 0 & 24  & 0 & 3 & -3 & 0 & -1 & 6 & 0 & 0\\
			$b$&  4& $(1,-2)\times (-5,-1)\times (-1,1)$ & -4 & -36 & - & - & -8 & 0 & -2 & -10 & 0 & 0 \\
			$c$&  4& $(5,-2)\times (-1,0)\times (-1,-1)$ & -3 & 3 & - & - & - & - & 0 & -2 & 0 & 0\\
			\hline
			1&   12& $(1,0)\times (1,0)\times (2,0)$& \multicolumn{10}{c|}{$X_A =\frac{102}{5} X_B =\frac{17}{24} X_C =51X_D$}\\
			2&   2& $(1,0)\times (0,-1)\times (0,2)$ & \multicolumn{10}{c|}{$\beta^g_1=-2, \beta^g_2=18$}\\
			&      &	                                                   & \multicolumn{10}{c|}{$\chi_1=\frac{\sqrt{\frac{85}{3}}}{4}, \chi_2=12\sqrt{\frac{51}{5}}, \chi_3=\sqrt{\frac{72}{15}}$}\\               
			\hline
		\end{tabular}
	\end{center}
\end{table}

\begin{table}[ht]
	\footnotesize
	\renewcommand{\arraystretch}{1.0}
	\caption{D6-brane configurations and intersection numbers of Model \ref{model6}, and its MSSM gauge coupling relation is $g^2_a=\frac{55}{597}g^2_b=\frac{7219}{5395}(\frac{5}{3}g^2_Y)=\frac{32\times 3^{1/4}85^{3/4}}{1393}\pi e^{\phi^4}$.} \label{model6}
	\begin{center}
		\begin{tabular}{|c||c|c||c|c|c|c|c|c|c|c|c|c|}
			\hline
			\rm{Model} \ref{model6} & \multicolumn{12}{c|}{$U(4)\times U(2)_L\times U(2)_R\times USp(12)\times USp(2)$}\\
			\hline \hline \rm{stack} & $N$ & $(n^1,l^1)\times (n^2,l^2)\times
			(n^3,l^3)$ & $n_{\Ysymm}$& $n_{\Yasymm}$ & $b$ & $b'$ & $c$ & $c'$& 1 & 2 & 3 & 4 \\
			\hline
			$a$&  8& $(-6,-1)\times (1,1)\times (1,1)$ & 0 & -24  & 3 & 0 & 0 & -3 & 1 & 0 & -6 & 0\\
			$b$&  4& $(-1,0)\times (-5,-2)\times (1,-1)$ & 3 & -3 & - & - & -8 & 0 & 0 & 0 & 2 & 0 \\
			$c$&  4& $(5,-1)\times (1,2)\times (1,1)$ & -3 & 3 & - & - & - & - & 2 & 0 & 10 & 0\\
			\hline
			1&   12& $(1,0)\times (1,0)\times (2,0)$& \multicolumn{10}{c|}{$X_A =\frac{17}{24} X_B =\frac{102}{5} X_C =51X_D$}\\
			3&   2& $(0,-1)\times (1,0)\times (0,2)$ & \multicolumn{10}{c|}{$\beta^g_1=-2, \beta^g_3=18$}\\
			&      &	                                                   & \multicolumn{10}{c|}{$\chi_1=12\sqrt{\frac{51}{5}}, \chi_2=\frac{\sqrt{\frac{85}{3}}}{4}, \chi_3=\sqrt{\frac{72}{15}}$}\\               
			\hline
		\end{tabular}
	\end{center}
\end{table}

\begin{table}[ht]
	\footnotesize
	\renewcommand{\arraystretch}{1.0}
	\caption{D6-brane configurations and intersection numbers of Model \ref{model3}, and its MSSM gauge coupling relation is $g^2_a=\frac{63}{260}g^2_b=\frac{575}{698}(\frac{5}{3}g^2_Y)=\frac{8\times 259^{3/4}}{65\sqrt{3}}\pi e^{\phi^4}$.} \label{model3}
	\begin{center}
		\begin{tabular}{|c||c|c||c|c|c|c|c|c|c|c|c|c|}
			\hline
			\rm{Model} \ref{model3} & \multicolumn{12}{c|}{$U(4)\times U(2)_L\times U(2)_R\times USp(2)\times USp(8)\times USp(6)$}\\
			\hline \hline \rm{stack} & $N$ & $(n^1,l^1)\times (n^2,l^2)\times
			(n^3,l^3)$ & $n_{\Ysymm}$& $n_{\Yasymm}$ & $b$ & $b'$ & $c$ & $c'$& 1 & 2 & 3 & 4 \\
			\hline
			$a$&  8& $(-1,0)\times (4,1)\times (-1,1)$ & 3 & -3  & 3 & 0 & 0 & -3 & 0 & 0 & 0 & -4\\
			$b$&  4& $(7,3)\times (-1,0)\times (1,1)$ & 10 & -10 & - & - & -1& 0  & 0 & -3 & 0 & -7\\
			$c$&  4& $(2,1)\times (1,-1)\times (1,-1)$ & 0 & -8 & - & - & - & - & -1 & 1 & 0 & -2 \\
			
			\hline
			1&   2& $(1,0)\times (1,0)\times (2,0)$& \multicolumn{10}{c|}{$X_A =\frac{37}{7} X_B =\frac{37}{12} X_C =\frac{37}{3}X_D$}\\
			2&   8& $(1,0)\times (0,-1)\times (0,2)$ & \multicolumn{10}{c|}{$\beta^g_1=-5, \beta^g_2=-2, \beta^g_4=11$}\\
			4 &  6 &	$(0,-1)\times (0,1)\times (2,0)$   & \multicolumn{10}{c|}{$\chi_1=\frac{\sqrt{259}}{6}, \chi_2=2\sqrt{\frac{37}{7}}, \chi_3=\sqrt{\frac{37}{5}}$}\\               
						\hline
		\end{tabular}
	\end{center}
\end{table}

\begin{table}[ht]
	\footnotesize
	\renewcommand{\arraystretch}{1.0}
	\caption{D6-brane configurations and intersection numbers of Model \ref{model4}, and its MSSM gauge coupling relation is $g^2_a=\frac{115}{156}g^2_b=\frac{105}{302}(\frac{5}{3}g^2_Y)=\frac{8\times 259^{3/4}}{65\sqrt{3}}\pi e^{\phi^4}$.} \label{model4}
	\begin{center}
		\begin{tabular}{|c||c|c||c|c|c|c|c|c|c|c|c|c|}
			\hline
			\rm{Model} \ref{model4} & \multicolumn{12}{c|}{$U(4)\times U(2)_L\times U(2)_R\times USp(2)\times USp(8)\times USp(6)$}\\
			\hline \hline \rm{stack} & $N$ & $(n^1,l^1)\times (n^2,l^2)\times
			(n^3,l^3)$ & $n_{\Ysymm}$& $n_{\Yasymm}$ & $b$ & $b'$ & $c$ & $c'$& 1 & 2 & 3 & 4 \\
			\hline
			$a$&  8& $(1,0)\times (-4,1)\times (-1,-1)$ & -3 & 3  & 0 & 3 & -3 & 0 & 0 & 0 & 0 & 4\\
			$b$&  4& $(2,-1)\times (-1,-1)\times (-1,-1)$ & 0 & 8 & - & - & -1& 0  & 1 & -1 & 0 & 2\\
			$c$&  4& $(-7,3)\times (1,0)\times (1,-1)$ & -10 & 10 & - & - & - & - & 0 & 3 & 0 & 7 \\
			
			\hline
			1&   2& $(1,0)\times (1,0)\times (2,0)$& \multicolumn{10}{c|}{$X_A =\frac{37}{7} X_B =\frac{37}{12} X_C =\frac{37}{3}X_D$}\\
			2&   8& $(1,0)\times (0,-1)\times (0,2)$ & \multicolumn{10}{c|}{$\beta^g_1=-5, \beta^g_2=-2, \beta^g_4=11$}\\
			4 &  6 &	$(0,-1)\times (0,1)\times (2,0)$   & \multicolumn{10}{c|}{$\chi_1=\frac{\sqrt{259}}{6}, \chi_2=2\sqrt{\frac{37}{7}}, \chi_3=\sqrt{\frac{37}{7}}$}\\               
						\hline
		\end{tabular}
	\end{center}
\end{table}

\begin{table}[ht]
	\footnotesize
	\renewcommand{\arraystretch}{1.0}
	\caption{D6-brane configurations and intersection numbers of Model \ref{model7}, and its MSSM gauge coupling relation is $g^2_a=\frac{2}{49}g^2_b=\frac{146}{71}(\frac{5}{3}g^2_Y)=\frac{80\times 2^{3/4}\sqrt{5}}{49\times 3^{1/4}}\pi e^{\phi^4}$.} \label{model7}
	\begin{center}
		\begin{tabular}{|c||c|c||c|c|c|c|c|c|c|c|c|c|}
			\hline
			\rm{Model} \ref{model7} & \multicolumn{12}{c|}{$U(4)\times U(2)_L\times U(2)_R\times USp(4)\times USp(2)$}\\
			\hline \hline \rm{stack} & $N$ & $(n^1,l^1)\times (n^2,l^2)\times
			(n^3,l^3)$ & $n_{\Ysymm}$& $n_{\Yasymm}$ & $b$ & $b'$ & $c$ & $c'$& 1 & 2 & 3 & 4 \\
			\hline
			$a$&  8& $(-1,0)\times (4,3)\times (-1,1)$ & 1 & -1  & 3 & 0 & 0 & -3 & 0 & 0 & 0 & -4\\
			$b$&  4& $(-8,-1)\times (-1,0)\times (-1,-1)$ & 9 & -9 & - & - & -20 & 0 & 0 & -1 & 0 & -8 \\
			$c$&  4& $(4,3)\times (-1,1)\times (-1,1)$ & -16 & -32 & - & - & - & - & 0 & 3 & 0 & -4\\
			\hline
			2&   4& $(1,0)\times (0,-1)\times (0,2)$& \multicolumn{10}{c|}{$X_A =\frac{25}{18} X_B =\frac{25}{3} X_C =\frac{100}{9}X_D$}\\
			4&   2& $(0,-1)\times (0,1)\times (0,2)$ & \multicolumn{10}{c|}{$\beta^g_2=-2, \beta^g_4=14$}\\
			&      &	                                                   & \multicolumn{10}{c|}{$\chi_1=10\sqrt{\frac{2}{3}}, \chi_2=\frac{\sqrt{5\frac{2}{3}}}{3}, \chi_3=\frac{5}{\sqrt{6}}$}\\               
				\hline
		\end{tabular}
	\end{center}
\end{table}

\begin{table}[ht]
	\footnotesize
	\renewcommand{\arraystretch}{1.0}
	\caption{D6-brane configurations and intersection numbers of Model \ref{model9}, and its MSSM gauge coupling relation is $g^2_a=\frac{379}{48}g^2_b=\frac{15}{118}(\frac{5}{3}g^2_Y)=\frac{3\times 3^{1/4}29^{3/4}}{7\sqrt{2}}\pi e^{\phi^4}$.} \label{model9}
	\begin{center}
		\begin{tabular}{|c||c|c||c|c|c|c|c|c|c|c|c|c|}
			\hline
			\rm{Model} \ref{model9} & \multicolumn{12}{c|}{$U(4)\times U(2)_L\times U(2)_R\times USp(6)\times USp(2)$}\\
			\hline \hline \rm{stack} & $N$ & $(n^1,l^1)\times (n^2,l^2)\times
			(n^3,l^3)$ & $n_{\Ysymm}$& $n_{\Yasymm}$ & $b$ & $b'$ & $c$ & $c'$& 1 & 2 & 3 & 4 \\
			\hline
			$a$&  8& $(1,0)\times (-4,3)\times (-1,-1)$ & -1 & 1  & 3 & 0 & -3 & 0 & 0 & 0 & 0 & 4\\
			$b$&  4& $(5,3)\times (-1,1)\times (-1,1)$ & -20 & -40 & - & - & 0 & 22 & 0 & 3 & 0 & -5 \\
			$c$&  4& $(-9,1)\times (-1,0)\times (-1,1)$ & -16 & -32 & - & - & - & - & 0 & 1 & 0 & 9\\
			\hline
			2&   6& $(1,0)\times (0,-1)\times (0,2)$& \multicolumn{10}{c|}{$X_A =\frac{116}{81} X_B =\frac{29}{3} X_C =\frac{116}{9}X_D$}\\
			4&   2& $(0,-1)\times (0,1)\times (0,2)$ & \multicolumn{10}{c|}{$\beta^g_2=-2, \beta^g_4=16$}\\
			&      &	                                                   & \multicolumn{10}{c|}{$\chi_1=\sqrt{87}, \chi_2=\frac{4\sqrt{\frac{29}{3}}}{9}, \chi_3=\frac{2\sqrt{\frac{29}{3}}}{3}$}\\               
					\hline
		\end{tabular}
	\end{center}
\end{table}

\begin{table}[ht]
	\footnotesize
	\renewcommand{\arraystretch}{1.0}
	\caption{D6-brane configurations and intersection numbers of Model \ref{model10}, and its MSSM gauge coupling relation is $g^2_a=\frac{9}{868}g^2_b=\frac{14395}{6874}(\frac{5}{3}g^2_Y)=\frac{24\times 3^{1/4}109^{3/4}}{217}\pi e^{\phi^4}$.} \label{model10}
	\begin{center}
		\begin{tabular}{|c||c|c||c|c|c|c|c|c|c|c|c|c|}
			\hline
			\rm{Model} \ref{model10} & \multicolumn{12}{c|}{$U(4)\times U(2)_L\times U(2)_R\times USp(2)\times USp(4)\times USp(2)$}\\
			\hline \hline \rm{stack} & $N$ & $(n^1,l^1)\times (n^2,l^2)\times
			(n^3,l^3)$ & $n_{\Ysymm}$& $n_{\Yasymm}$ & $b$ & $b'$ & $c$ & $c'$& 1 & 2 & 3 & 4 \\
			\hline
			$a$&  8& $(-1,0)\times (4,3)\times (-1,1)$ & 1 & -1  & 3 & 0 & -3 & 0 & 0 & 0 & 0 & -4\\
			$b$&  4& $(9,1)\times (-1,0)\times (1,1)$ & 1 & -10 & - & - & 0& -23  & 0 & -1 & 0 & -9\\
			$c$&  4& $(-4,3)\times (1,1)\times (-1,-1)$ & -20 & -40 & - & - & - & - & 3 & -3 & 0 & 4 \\
			
			\hline
			1&   2& $(1,0)\times (1,0)\times (2,0)$& \multicolumn{10}{c|}{$X_A =\frac{109}{81} X_B =\frac{109}{12} X_C =\frac{109}{9}X_D$}\\
			2&   4& $(1,0)\times (0,-1)\times (0,2)$ & \multicolumn{10}{c|}{$\beta^g_1=-3, \beta^g_2=-2, \beta^g_4=15$}\\
			4 &  2 &	$(0,-1)\times (0,1)\times (2,0)$   & \multicolumn{10}{c|}{$\chi_1=\frac{\sqrt{327}}{2}, \chi_2=\frac{2\sqrt{\frac{109}{3}}}{9}, \chi_3=\frac{\sqrt{\frac{109}{3}}}{3}$}\\               
						\hline
		\end{tabular}
	\end{center}
\end{table}

\begin{table}[ht]
	\footnotesize
	\renewcommand{\arraystretch}{1.0}
	\caption{D6-brane configurations and intersection numbers of Model \ref{model8}, and its MSSM gauge coupling relation is $g^2_a=\frac{5}{98}g^2_b=\frac{1825}{856}(\frac{5}{3}g^2_Y)=\frac{200\times 2^{3/4}}{49\times 3^{1/4}}\pi e^{\phi^4}$.} \label{model8}
	\begin{center}
		\begin{tabular}{|c||c|c||c|c|c|c|c|c|c|c|c|c|}
			\hline
			\rm{Model} \ref{model8} & \multicolumn{12}{c|}{$U(4)\times U(2)_L\times U(2)_R\times USp(2)\times USp(6)\times USp(2)$}\\
			\hline \hline \rm{stack} & $N$ & $(n^1,l^1)\times (n^2,l^2)\times
			(n^3,l^3)$ & $n_{\Ysymm}$& $n_{\Yasymm}$ & $b$ & $b'$ & $c$ & $c'$& 1 & 2 & 3 & 4 \\
			\hline
			$a$&  8& $(-1,0)\times (4,3)\times (-1,1)$ & 1 & -1  & 3 & 0 & 0 & -3 & 0 & 0 & 0 & -4\\
			$b$&  4& $(-10,-1)\times (1,0)\times (1,1)$ & 11 & -11 & - & - & -25& 0  & 0 & -1 & 0 & -10\\
			$c$&  4& $(5,3)\times (-1,1)\times (-1,1)$ & -20 & -40 & - & - & - & - & -3 & 3 & 0 & -5 \\			
			\hline
			1&   2& $(1,0)\times (1,0)\times (2,0)$& \multicolumn{10}{c|}{$X_A =\frac{25}{18} X_B =\frac{125}{12} X_C =\frac{125}{9}X_D$}\\
			2&   6& $(1,0)\times (0,-1)\times (0,2)$ & \multicolumn{10}{c|}{$\beta^g_1=-3, \beta^g_2=-2, \beta^g_4=17$}\\
			4 &  2 &	$(0,-1)\times (0,1)\times (2,0)$   & \multicolumn{10}{c|}{$\chi_1=\frac{25}{\sqrt{6}}, \chi_2=\frac{5\sqrt{\frac{2}{3}}}{3}, \chi_3=\frac{5}{\sqrt{6}}$}\\               		
			\hline
		\end{tabular}
	\end{center}
\end{table}

\begin{table}[ht]
	\footnotesize
	\renewcommand{\arraystretch}{1.0}
	\caption{D6-brane configurations and intersection numbers of Model \ref{model1}, and its MSSM gauge coupling relation is $g^2_a=\frac{15}{7}g^2_b=\frac{5}{11}(\frac{5}{3}g^2_Y)=\frac{16\times 2^{1/4}5^{3/4}}{7\sqrt{3}}\pi e^{\phi^4}$.} \label{model1}
	\begin{center}
		\begin{tabular}{|c||c|c||c|c|c|c|c|c|c|c|c|c|}
			\hline
			\rm{Model} \ref{model1} & \multicolumn{12}{c|}{$U(4)\times U(2)_L\times U(2)_R\times USp(4)\times USp(6)$}\\
			\hline \hline \rm{stack} & $N$ & $(n^1,l^1)\times (n^2,l^2)\times
			(n^3,l^3)$ & $n_{\Ysymm}$& $n_{\Yasymm}$ & $b$ & $b'$ & $c$ & $c'$& 1 & 2 & 3 & 4 \\
			\hline
			$a$&  8& $(1,0)\times (2,-1)\times (1,1)$ & -1 & 1  & 0 & 3 & -3 & 0 & 0 & 0 & -1 & 2\\
			$b$&  4& $(10,3)\times (-1,0)\times (-1,-1)$ & -13 & -13 & - & - & 16 & 0 & 0 & 0 & 0 & 10 \\
			$c$&  4& $(-2,1)\times (5,-1)\times (1,-1)$ & 4 & 36 & - & - & - & - & 0 & 0 & 2 & 10\\
			\hline
			3&   4& $(0,-1)\times (1,0)\times (0,2)$& \multicolumn{10}{c|}{$X_A =\frac{4}{5} X_B =\frac{4}{3} X_C =\frac{8}{3}X_D$}\\
			4&   6& $(0,-1)\times (0,1)\times (2,0)$ & \multicolumn{10}{c|}{$\beta^g_3=-2, \beta^g_4=18$}\\
			&      &	                                                   & \multicolumn{10}{c|}{$\chi_1=\frac{2\sqrt{10}}{3}, \chi_2=2\sqrt{\frac{2}{5}}, \chi_3=2\sqrt{\frac{2}{5}}$}\\               			
			\hline
		\end{tabular}
	\end{center}
\end{table}

\begin{table}[ht]
	\footnotesize
	\renewcommand{\arraystretch}{1.0}
	\caption{D6-brane configurations and intersection numbers of Model \ref{model2}, and its MSSM gauge coupling relation is $g^2_a=\frac{1}{3}g^2_b=\frac{25}{17}(\frac{5}{3}g^2_Y)=\frac{16\times 2^{1/4}5^{3/4}}{7\sqrt{3}}\pi e^{\phi^4}$.} \label{model2}
	\begin{center}
		\begin{tabular}{|c||c|c||c|c|c|c|c|c|c|c|c|c|}
			\hline
			\rm{Model} \ref{model2} & \multicolumn{12}{c|}{$U(4)\times U(2)_L\times U(2)_R\times USp(4)\times USp(6)$}\\
			\hline \hline \rm{stack} & $N$ & $(n^1,l^1)\times (n^2,l^2)\times
			(n^3,l^3)$ & $n_{\Ysymm}$& $n_{\Yasymm}$ & $b$ & $b'$ & $c$ & $c'$& 1 & 2 & 3 & 4 \\
			\hline
			$a$&  8& $(-1,0)\times (2,1)\times (-1,1)$ & 1 & -1  & 3 & 0 & 0 & -3 & 0 & 0 & 1 & -2\\
			$b$&  4& $(-2,-1)\times (5,1)\times (1,1)$ & -4 & -36 & - & - & 16& 0  & 0 & 0 & -2 & -10\\
			$c$&  4& $(-10,3)\times (1,0)\times (-1,1)$ & 13 & -13 & - & - & - & - & 0 & 0 & 0 & -10 \\
						\hline
			3&   4& $(0,-1)\times (1,0)\times (0,2)$& \multicolumn{10}{c|}{$X_A =\frac{4}{5} X_B =\frac{4}{3} X_C =\frac{8}{3}X_D$}\\
			4&   6& $(0,-1)\times (0,1)\times (2,0)$ & \multicolumn{10}{c|}{$\beta^g_3=-2, \beta^g_4=18$}\\
			&      &	                                                   & \multicolumn{10}{c|}{$\chi_1=\frac{2\sqrt{10}}{3}, \chi_2=2\sqrt{\frac{2}{5}}, \chi_3=2\sqrt{\frac{2}{5}}$}\\               
					\hline
		\end{tabular}
	\end{center}
\end{table}

\end{appendix}

\end{document}